\documentclass[12pt,a4paper,twoside]{article}
\usepackage{amsmath}
\usepackage{amssymb}   
\usepackage{amsfonts}
\usepackage{theorem}
\usepackage{epsfig}
\usepackage{helvet}

\newcommand{\R}{{\mathord{\mathbb R}}}
\newcommand{\Z}{{\mathord{\mathbb Z}}}
\newcommand{\N}{{\mathord{\mathbb N}}}
\newcommand{\C}{{\mathord{\mathbb C}}}

\newcommand{\T}{{\mathord{\mathbb T}}}

\newcommand{\mB}{{\mathcal B}}

\newcommand{\mE}{{\mathcal E}}

\newcommand{\mG}{{\mathcal G}}
\newcommand{\mH}{{\mathcal H}}

\newcommand{\mL}{{\mathcal L}}

\newcommand{\mO}{{\mathcal O}}

\newcommand{\mR}{{\mathcal R}}
\newcommand{\mS}{{\mathcal S}}

\newcommand{\mU}{{\mathcal U}}

\renewcommand{\Re}{{\rm Re}}
\renewcommand{\Im}{{\rm Im}}
\newcommand{\e}{{\rm e}}

\newcommand{\dG}{{\rm d}\Gamma}

\newcommand{\rpp}{{\rm pp}}

\newcommand{\rsc}{{\rm sc}}

\renewcommand{\i}{{\rm i}}

\newcommand{\ran}{{\rm ran\,}}
\newcommand{\sign}{{\rm sign}}
\newcommand{\tr}{{\rm tr}}
\newcommand{\res}{{\rm res}}
\def\slim{\mathop{\rm s-lim}}
\def\wlim{\mathop{\rm w-lim}}
\newcommand{\Aut}{{\rm Aut}}

\newcommand{\rd}{{\rm d}}

\newcommand{\diag}{{\rm diag}}
\newcommand{\sh}{{\rm sh}}

\newcommand{\ch}{{\rm ch}}
\newcommand{\fh}{{{\mathfrak h}}}

\newcommand{\ff}{{{\mathfrak f}}}

\newcommand{\fs}{{\mathfrak s}}
\newcommand{\fF}{{\mathfrak F}}
\newcommand{\fA}{{\mathfrak A}}

\newcommand{\is}{i_\mS^{}}
\newcommand{\iss}{i_\mS^\ast}
\newcommand{\ir}{i_\mR^{}}
\newcommand{\irs}{i_\mR^\ast}

\newcommand{\iLs}{i_L^\ast}

\newcommand{\iRs}{i_R^\ast}
\newcommand{\ia}{i_a^{}}
\newcommand{\ias}{i_a^\ast}
\newcommand{\planck}{\varrho^{}}
\newcommand{\density}{\rho^{}}
\newcommand{\lam}{\lambda}
\newcommand{\kp}{\kappa}
\newcommand{\veps}{{\varepsilon}}
\newcommand{\eps}{{\epsilon}}

\newcommand{\spec}{{\rm spec}}

\newcommand{\Ep}{{\rm Ep}}
\newcommand{\gam}{\gamma}
\newcommand{\sg}{\sigma}

\newcommand{\wt}{\widetilde}
\newcommand{\wh}{\widehat}
\newtheorem{thm}{Theorem}
\newtheorem{proposition}[thm]{Proposition}
\newtheorem{lemma}[thm]{Lemma}
\newtheorem{definition}[thm]{Definition}
{\theorembodyfont{\upshape} \newtheorem{remark}[thm]{\it Remark}}

\newtheorem{corollary}[thm]{Corollary}
\newcommand{\bd}{\begin{definition}}
\newcommand{\ed}{\end{definition}\vspace{1mm}}
\newcommand{\bt}{\begin{thm}}\vspace{1mm}
\newcommand{\et}{\end{thm}}
\newcommand{\bc}{\begin{corollary}}
\newcommand{\ec}{\end{corollary}\vspace{1mm}}
\newcommand{\bl}{\begin{lemma}}
\newcommand{\el}{\end{lemma}\vspace{1mm}}
\newcommand{\bp}{\begin{proposition}}
\newcommand{\ep}{\end{proposition}\vspace{1mm}}
\newcommand{\br}{\begin{remark}}
\newcommand{\er}{\end{remark}}
\newcommand{\bprf}{\noindent{\it Proof.}\, }
\newcommand{\eprf}{\hfill $\Box$ \vspace{5mm}}

\def\bas#1\eas{\begin{align*}#1\end{align*}}
\def\ba#1\ea{\begin{align}#1\end{align}}

\begin{document}
\pagestyle{myheadings}
\markboth{W. H. Aschbacher}{From the microscopic to the van Hove
regime}

\title{From the microscopic to the van Hove regime in the XY chain out 
of equilibrium}

\author{Walter H. Aschbacher\footnote{walter.aschbacher@cpt.univ-mrs.fr}
\\ \\
Aix Marseille Universit\'e, CNRS, CPT, UMR 7332, 13288 Marseille, France\\
Universit\'e de Toulon, CNRS, CPT, UMR 7332, 83957 La Garde, France 
}

\date{}
\maketitle
\begin{abstract}
Using the framework of rigorous algebraic quantum statistical mechanics, 
we construct the unique nonequilibrium steady state 
in the isotropic XY chain in which a sample of arbitrary finite size
is coupled by a bond coupling perturbation of arbitrary strength to two infinitely extended thermal reservoirs, and we prove that this state is thermodynamically nontrivial. Moreover, extracting the leading second 
order contribution to its microscopic entropy production and deriving 
its entropy production in the van Hove weak coupling regime, we prove 
that, in the mathematically and physically important XY chain, the van Hove regime reproduces the leading order contribution to the microscopic regime. 
\end{abstract}

\section{Introduction}

In recent years, a broad range of important thermodynamic properties 
of open quantum systems have been successfully derived
from first principles within the mathematically rigorous framework of algebraic quantum statistical mechanics. Not only return to equilibrium
type phenomena from states close to equilibrium have been explored
but also fundamental transport processes in systems far from equilibrium  have come within reach. 
In the latter field, an important role is played by the quasifree fermionic systems since, on one hand, they allow for a powerful description by means of scattering theory on the one-particle Hilbert space on which the fermionic algebra of observables is built
being thus ideally suited for a rigorous analysis on many levels. On the other hand, they also constitute  a class of systems which are indeed realized in nature. One of the most prominent representatives of this
class is the XY spin chain introduced mathematically in 1961 by Lieb 
{\it et al.} \cite{LSM} who showed that this spin system 
can be mapped onto a gas of free fermions by using the Jordan-Wigner transformation. Already at the end of the 1960s,
the first candidates for a possible physical realization have been identified
by Culvahouse {\it et al.} \cite{CSP} and, later, by D'Iorio {\it et al.} \cite{DAT} (see also Sologubenko {\it et al.} \cite{SGOVR} for  experiments on more general Heisenberg models). Subsequently, 
Araki \cite{A3} extended the mathematical setup from the finite spin chain to fermions over the two-sided infinite discrete line in the framework of $C^\ast$-dynamical systems and it is this 
system whose energy transport properties we will study in this paper. 
In order to do so, 
we fall back upon the paradigm of the theory of open system by 
coupling a localized sample to two infinitely extended reservoirs in thermal equilibrium at different temperatures. For this purpose, we cut the
two bonds between the sites $\pm n$ and $\pm(n+1)$ of the two-sided discrete line
meaning  that the coupling strength in the local Hamiltonian between the corresponding sites, initially at value $\lam=1$,
is set to zero.
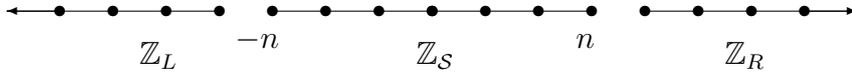
\begin{figure}
\setlength{\unitlength}{7mm}
\begin{picture}(-1,1)
\put(3,0){\line(1,0){4}}
\put(8,0){\line(1,0){6}}
\put(15,0){\line(1,0){4}}
\multiput(4,0)(1,0){4}{\circle*{0.2}}
\multiput(8,0)(1,0){7}{\circle*{0.2}}
\put(11,0){\circle*{0.2}}
\multiput(15,0)(1,0){4}{\circle*{0.2}}
\put(13.7,-0.7){$n$}
\put(7.3,-0.7){$-n$}
\put(18,0){\vector(1,0){1}}
\put(4,0){\vector(-1,0){1}}
\put(10.7,-1){$\Z_\mS$}
\put(16.5,-1){$\Z_R$}
\put(5.5,-1){$\Z_L$}
\end{picture}
\vspace{6mm}
\caption{The nonequilibrium setting for the XY chain.}
\label{fig:nonequilibrium}
\end{figure}
The piece $\Z_\mS$ between these bonds plays the role of the configuration space of the sample whereas the remaining two half-infinite pieces $\Z_L$ and $\Z_R$ to its left and right constitute the configuration spaces of the reservoirs, see Figure \ref{fig:nonequilibrium}.
Over these configuration spaces, an initial state is prepared as the product of three thermal equilibrium states. The first central object of interest is then the so-called nonequilibrium steady state (NESS) defined 
by Ruelle \cite{R1} as the large time limit of the (averaged) trajectory of the initial state along the  fully coupled time evolution.
In this setup, the unique NESS has been constructed
in Aschbacher and Pillet \cite{AP} using Ruelle's scattering approach (for a special case of the XY model, namely for the case of vanishing anisotropy and external magnetic field parameters given in Remark \ref{rem:XY} below, this NESS has also been found by Araki and Ho \cite{AH} using a different method).
It has been proved in \cite{AP} that this NESS, henceforth called the XY NESS, 
is thermodynamically nontrivial in the sense that its entropy production  is strictly positive as soon as the system is truly out of equilibrium. 
In the present paper, we generalize the foregoing situation to couplings
of arbitrary strength $\lam\in\R$. As a first result, we prove that the XY NESS can be embedded into a two-parameter family of NESS parametrized by  $\lam$ and  $n$ which, for 
$\lam\neq 0$, are all
thermodynamically nontrivial. This provides us with a physically richer
nonequilibrium situation. In particular,  it becomes possible to study the van Hove weak
coupling regime $\lam\to 0$ of the entropy production of these NESS
and compare it to the leading order contribution of the fully
microscopic regime.
Although one naturally expects that the van Hove regime reproduces the leading order contribution to the microscopic regime, this has been
proven for few systems only. One of these systems is the so-called
simple electronic black box model from Aschbacher {\it et al.} 
\cite{AJPP1} (which corresponds to $n=0$) and another one is the spin-fermion system from Jak\v si\'c {\it et al.} \cite{JOP}. Then, as a second result, and this is the main motivation of the present
paper, we prove that this natural expectation is indeed rigorously true
in the mathematically and physically important XY chain out of 
equilibrium.

The present paper is organized as follows. In Section \ref{sec:mi}, we introduce
the nonequilibrium setting, construct a (family of) unique NESS, and derive an explicit expression for its entropy production. In particular, we prove that, in a true nonequilibrium, this NESS is 
thermodynamically nontrivial for all nonvanishing couplings and all sample sizes. In Section \ref{sec:macro}, we extract the leading 
second order contribution to the microscopic entropy production. 
In Section \ref{sec:vanHove}, we construct the NESS in the van Hove regime and prove that the van Hove entropy production  is the leading order contribution to the microscopic entropy production.
In Appendix \ref{app:spec}, some spectral properties 
of the appearing one-particle Hamiltonians are summarized.
In Appendix \ref{app:wave}, we construct the wave operator needed in the
derivation of the microscopic NESS and display some results of the 
lengthy computations involved in its construction. Finally, in
Appendix \ref{app:davies}, we summarize the van Hove weak coupling
theory and derive the necessary decay and positivity properties  of the reservoir time correlation function.

\section{Microscopic regime}
\label{sec:mi} 

We begin this section by summarizing the setting for the system out of
equilibrium used in Aschbacher and Pillet \cite{AP}. In contradistinction
to the presentation there, we skip the formulation of the two-sided XY
chain as a spin system and rather focus directly on the underlying 
$C^\ast$-dynamical system structure in terms of Bogoliubov automorphisms 
on the CAR algebra of observables $\mO$ over the corresponding one-particle Hilbert space  $\fh$. 
Recall that a  $C^\ast$-dynamical system is a pair $(\mO,\tau)$ with
\ba
\R\ni t\mapsto \tau^t\in {\rm Aut}(\mO),
\ea
where $\mO$ is a $C^\ast$ algebra and 
$\tau^t$ a strongly continuous group of 
$\ast$-automorphism of $\mO$ (for more information on the algebraic approach to open quantum systems, see, for example, Aschbacher {\it et al.} \cite{AJPP1}). Moreover, let us denote the states on $\mO$ by
$\mE(\mO)$ and recall that a state $\omega\in\mE(\mO)$ 
is called a (gauge invariant) quasifree state with density 
$\rho\in\mL(\fh)$ satisfying $0\le\rho\le 1$ 
if, for all $p,q\in\N$ and all  $f_i, g_j\in\fh$ with $i,j\in\N$, we have
\ba
\omega(a^\ast(f_p)\ldots a^\ast(f_1)a(g_1)\ldots a(g_q))
=\delta_{pq}\det([(g_i,\rho f_j)]_{i,j=1}^p),
\ea
where $a^\ast(f), a(f)\in\mL(\fF(\fh))$ with $f\in\fh$ stand for the usual creation and annihilation operators on the fermionic Fock space 
$\fF(\fh)$ over the one-particle Hilbert space  $\fh$. Here, we used the 
notation $\mL(\mG,\mH)$ for the bounded linear operators
from some Hilbert space $\mG$ into some Hilbert space $\mH$ (with 
$\mL(\mH):=\mL(\mH,\mH)$), and $\delta_{ab}$ denotes the 
usual Kronecker symbol. 
In the following, we will also use the notations 
$\N_0:=\{0\}\cup\N$ and $\Re(A):=(A+A^\ast)/2$ and $\Im(A):=(A-A^\ast)/(2\i)$ for all $A\in\mL(\mH)$, and $\mL^0(\mH)$ 
and $\mL^1(\mH)$ stand for the finite rank operators and the trace class
operators, respectively. Moreover, $\rd\Gamma(A)$ is the usual second quantization on the fermionic Fock space.

\bd[Quasifree setting]
\label{def:qfs}
The ingredients for this setting are specified as follows.
\begin{enumerate}
\item[\textnormal{(a)}] \textnormal{Observable algebra}\\
Let $n\in\N_0$. The sample and the reservoir configuration spaces are defined by
\ba
\Z_\mS
&:=\{x\in\Z\,|\,\, |x|\le n\},\\
\Z_\mR
&:=\{x\in\Z\,|\,\, |x|\ge n+1\},
\ea
whereas the subreservoir spaces are given by 
\ba
\Z_L
&:=\{x\in\Z\,|\, x\le -(n+1)\},\\
\Z_R
&:=\{x\in\Z\,|\, x\ge n+1\},
\ea
see Figure \ref{fig:nonequilibrium}. The observable algebra is then defined to be the CAR algebra 
\ba
\mO
:=\fA(\fh_\mS\oplus\fh_\mR),
\ea
where the one-particle Hilbert space consists of the sample and the reservoir space
\ba
\fh_\mS
&:=\ell^2(\Z_\mS),\\
\fh_\mR
&:=\ell^2(\Z_\mR),
\ea
and the dimension of the sample Hilbert space is denoted by 
$n_\mS:=\dim(\fh_\mS)=2n+1$.
Moreover, the one-particle Hilbert spaces of the  subreservoirs are
given by
\ba
\fh_L
&:=\ell^2(\Z_L),\\
\fh_R
&:=\ell^2(\Z_R).
\ea
For $a\in\{\mS,\mR,L,R\}$, using the map $\ia\in\mL(\fh_a,\fh)$ defined, for all $f\in\fh_a$, by $\ia f(x):= f(x)$ if $x\in\Z_a$ and 
$\ia f(x):=0$ if $x\in\Z\setminus\Z_a$, the total one-particle Hilbert space is naturally identified with 
\ba
\fh
:=\ell^2(\Z)
\ea
through $f\mapsto\iss f\oplus \irs f$ for all
$f\in\fh$. Analogously, $\fh_\mR$ is identified with $\fh_L\oplus\fh_R$
through $f\mapsto\iLs\ir f\oplus \iRs\ir f$ for all 
$f\in\fh_\mR$.

\item[\textnormal{(b)}] \textnormal{Dynamics}\\
Let $\lam\in\R$. The one-particle Hamiltonians $h,h_\lam
\in\mL(\fh)$ are defined by
\ba
\label{h}
h
&:=
{\rm Re}[u],\\
\label{h0}
h_\lam
&:= h+(\lam-1)v\nonumber\\
&\hspace{1mm}=h_0+\lam v,
\ea
where the right translation $u\in\mL(\fh)$ is  defined by 
$(uf)(x):=f(x-1)$ for all $f\in\fh$ and all $x\in\Z$. 
The operator $h_1=h$ is 
called the XY Hamiltonian, $h_0$ the decoupled Hamiltonian,
and $h_\lam$ with $\lambda\neq 0$ the coupled Hamiltonian.
Moreover, the operator $v\in \mL^0(\fh)$, called the bond perturbation, is defined by
\ba
\label{v}
v
&:=\sum_{a\in\{L,R\}} v_a,\\
\label{va}
v_a
&:=\Re[(\delta_{\mS,a},\cdot\,)\, \delta_{\mR,a}],
\ea
where the coupling functions are specified by
\ba
\delta_{\mS,L}
:=\delta_{-n},\quad
\delta_{\mS,R}
:=\delta_{n},\quad
\delta_{\mR,L}
:=\delta_{-(n+1)},\quad
\delta_{\mR,R}
:=\delta_{n+1},
\ea
and $\delta_x\in\fh$ with $x\in\Z$ is
given by $\delta_x(y):=\delta_{xy}$ for all $y\in\Z$. Moreover,
for $a\in\{\mS,\mR,L,R\}$, the one-particle Hamiltonians 
$h_a\in\mL(\fh_a)$  are defined by
\ba
h_a
&:=\ias h\ia.
\ea
The second quantized Hamiltonians on the fermionic Fock space 
$\fF(\fh)$ are given by 
\ba
H
&:=\dG(h),\\
V
&:=\sum_{a\in\{L,R\}} V_a,\\
V_a
&:=\dG(v_a),\\
H_\lam
&:=H+(\lam-1)V\nonumber\\
&\hspace{1mm}=H_0+\lam V,
\ea
where $H$ is unbounded and $V_a\in \mO$ is a local perturbation.
Finally, the dynamics $\tau_\lam^t\in\Aut(\mO)$ with 
$t,\lam\in\R$ are defined, for all $A\in\mO$, by
\ba
\label{tau-lt}
\tau_\lam^t(A)
:=\e^{\i t H_\lam}\! A\, \e^{-\i t H_\lam}.
\ea

\item[\textnormal{(c)}] \textnormal{Initial state}\\
The initial state $\omega\in\mE(\mO)$ is defined to be 
quasifree
with
density $\density\in\mL(\fh)$ defined by
\ba
\label{rho}
\density
:=\is\density_\mS\iss
+\ir\density_\mR\irs,
\ea
where $\density_\mS\in\mL(\fh_\mS)$ and $\density_\mR\in\mL(\fh_\mR)$ are
given by
\ba
\label{rho-S}
\density_\mS
&:=\planck_{\beta_\mS}(h_\mS),\\
\label{rho-Res}
\density_\mR
&:=\sum_{a\in\{L,R\}} \irs\ia\planck_{\beta_a}(h_a)\ias\ir.
\ea
Here, for any $\alpha\in\R$, the Planck density function 
$\planck_\alpha: \R\to\R$ is defined by
\ba
\label{planck}
\varrho_\alpha(e)
:=\left(1+\e^{\alpha e}\right)^{-1},
\ea
{\it Pour fixer les id\'ees}, we will always assume 
that the inverse temperatures of the sample and the  reservoirs satisfy
\ba
\label{betas}
\beta_\mS
=0,\quad 
0
<\beta_L
\le\beta_R
<\infty,
\ea
and we set $\beta:=(\beta_R+\beta_L)/2$ and $\delta:=(\beta_R-\beta_L)/2$.
\end{enumerate}
\ed
\br
\label{rem:XY}
As discussed in the introduction, 
this model has its origin in the XY spin chain whose formal Hamiltonian
reads
\ba
\label{H-XY}
H_{XY}
=-\frac{1}{4}\sum_{x\in\Z}\left\{(1+\gamma)\,\sigma_1^{(x)}
\sigma_1^{(x+1)}+(1-\gamma)\,\sigma_2^{(x)}\sigma_2^{(x+1)}
+2\mu  \, \sigma_3^{(x)}\right\},
\ea
where $\gamma\in(-1,1)$ denotes the anisotropy, $\mu\in\R$ the external magnetic field, and the Pauli basis of $\C^{2\times 2}$ is given by
\ba
\label{Pauli} 
\sigma_0
=\left[\begin{array}{cc}
1 & 0
\\ 0&1
\end{array}\right],\quad 
\sigma_1
=\left[\begin{array}{cc}
0 & 1\\
1& 0
\end{array}\right],\quad 
\sigma_2
=\left[\begin{array}{cc}
0 &-\i\\ 
\i & 0
\end{array}\right],\quad
\sigma_3
=\left[\begin{array}{cc}
1 & 0
\\ 0& -1
\end{array}\right].
\ea
Namely, under the Araki-Jordan-Wigner transformation (see, for example,
Araki \cite{A3}), the Hamiltonian from \eqref{h} corresponds to the case of the 
so-called isotropic XY chain (or XX chain) without external magnetic
field, {\it i.e.}  to the case where $\gamma=0$ and $\mu=0$.
In order to treat the anisotropic case $\gam\neq 0$, one often uses the
so-called selfdual quasifree setup introduced and developed in Araki
\cite{A1, A2}. There, one works in the doubled one-particle Hilbert space
$\fh^{\oplus 2}$ and the generator of the truly anisotropic XY dynamics 
has nontrivial off-diagonal blocks on $\fh^{\oplus 2}$ (which vanish
for $\gam=0$). In many respects, the truly anisotropic XY model is substantially more complicated than the isotropic one (this is true
{\it a fortiori} if a magnetic field is added whose contribution to the generator acts diagonally on $\fh^{\oplus 2}$ though). In the following,
every once in a while, we will make a remark on the corresponding
issue for the anisotropic case.
\er
\br
\label{rem:dec}
The Hamiltonian $h_\lam$ for $\lam=0$ does not couple the different subsystems to each other, {\it i.e.} we have 
$h_0=\sum_{a\in\{\mS,L,R\}} \ia h_a\ias$.
\er

In order to construct a NESS, we use the following definition due to Ruelle \cite{R1}. If not specified otherwise, it will always be 
assumed that $n\in\N_0$ and $\lam\in\R$.
\bd[NESS]
\label{def:ness}
A NESS associated with the $C^\ast$-dynamical system 
$(\mO,\tau_\lambda^t)$ and the initial state $\omega\in\mE(\mO)$ is a weak-$\ast$ limit point for $T\to\infty$ of the net
\ba
\label{ness-0}
\left\{\frac1T\int_0^T\rd t\,\, \omega\circ \tau_\lambda^t 
\,\,\Big|\,\,T>0\right\}.
\ea
Such a NESS is denoted by $\omega_{\lam,+}\in\mE(\mO)$.
\ed

If the coupled time evolution is the XY dynamics ({\it i.e.} if $\lam=1$),
the unique quasifree NESS in the fully anisotropic XY model with magnetic field has been constructed in Aschbacher and Pillet \cite{AP}. In order to state
the corresponding theorem, we switch to the momentum space representation
by using  the Fourier transformation $\ff:\fh\to\widehat\fh$, 
where
\ba
\label{h-hat}
\widehat\fh
:=L^2([-\pi,\pi];\tfrac{\rd k}{2\pi}),
\ea
and $\ff$ is defined with the sign convention $\widehat f(k):=(\ff f) (k):=
\sum_{x\in\Z}f(x)\e^{\i k x}$. Moreover, we also use the notation 
$\widehat A:=\ff A \ff^\ast$ for all $A\in\mL(\fh)$. In the following,
for any selfadjoint 
$A,B\in\mL(\fh)$, we denote by $w(A,B)\in\mL(\fh)$ the wave operator
\ba
\label{waveAB-0}
w(A,B)
:=\slim_{t\to\infty} {\rm e}^{-{\rm i}t A} 
{\rm e}^{{\rm i}tB}1_{\rm ac}(B),
\ea
where $1_{\rm ac}(B)$ is the spectral projection onto the absolutely continuous subspace of $B$. 
\bt[XY NESS]
\label{thm:ness}
There exists a unique quasifree NESS 
$\omega_{1,+}\in \mE(\mO)$ 
associated with the $C^\ast$-dynamical system $(\mO,\tau_1^t)$ and the
initial state $\omega\in\mE(\mO)$. Moreover, its density $\rho_{1,+}\in\mL(\fh)$ has the form
\ba
\rho_{1,+}
\label{rho1+-1}
&= w^\ast(h_0,h)\rho w(h_0,h)\nonumber\\
&=\big(1+\e^{\beta h-\delta d}\big)^{\!-1},
\ea
where, in momentum space, $\wh h, \wh d \in \mL(\wh\fh)$ are the multiplication operators acting, for all $\varphi\in \wh\fh$ and all 
$k\in (-\pi,\pi]$, as
\ba
\label{hath}
\wh h\varphi(k)
&=\eps(k)\varphi(k),\\
\label{hatr}
\wh d\varphi(k)
&=\sign(\eps'(k)) \,\wh h\varphi(k),
\ea
and the dispersion relation $\eps:\R\to\R$ is given by 
\ba
\eps(k)
:=\cos(k).
\ea
\et
\bprf
See Aschbacher and Pillet \cite{AP}.
\eprf

In the following, we denote by $1_e(h_\lambda)\in\mL(\fh)$ the usual spectral projection onto the eigenspace  of $h_\lambda $ corresponding to the eigenvalue $e\in\spec_\rpp(h_\lambda )$. Moreover, 
$\spec_\rsc(h_\lambda )$ is the singular continuous spectrum of $h_\lam$.
We then have the following result.
\bt[Microscopic NESS]
\label{thm:ness-2}
There exists a unique NESS $\omega_{\lambda,+}\in\mE(\mO)$ associated with the $C^\ast$-dynamical system $(\mO,\tau_\lambda^t)$ and the initial state $\omega\in\mE(\mO)$. Moreover, its density $\rho_{\lambda,+}\in\mL(\fh)$ has the form
\ba
\label{Skp}
\rho_{\lambda,+} 
=w^\ast(h_0,h_\lambda )\rho w(h_0,h_\lambda)
+\sum_{e\in\, \spec_\rpp(h_\lambda )} 1_e(h_\lambda)\rho 1_e(h_\lambda).
\ea
\et
\bprf
Since $h_\lam\in\mL(\fh)$, $h_\lam-h_0\in\mL^0(\fh)$, and $\spec_\rsc(h_\lam)=\emptyset$ due to Lemma \ref{lem:coupled} of Appendix \ref{app:spec}, we can use Aschbacher {\it et al.} \cite{AJPP2}
which yields the assertion. 
\eprf
\br
As given in Lemma \ref{lem:coupled} of Appendix \ref{app:spec}, the pure
point component in \eqref{Skp} is absent if $0<|\lam|\le 1$ (see also
Jak\v si\'c {\it et al.} \cite{JKP}).
\er

We next turn to the energy current observable and its NESS expectation.
\bd[Energy current]
\label{def:current}
The observable $\Phi_{\lam,a}\in\mO$ with $a\in\{L,R\}$
describing the energy current flowing from reservoir $a$ 
into the sample is defined by
\ba
\label{eco}
\Phi_{\lam,a}
:=\rd\Gamma(\varphi_{\lam,a}),
\ea
where the one-particle energy current observable 
$\varphi_{\lam,a}^{}\in\mL^0(\fh)$ is given by
\ba
\label{1eco-0}
\varphi_{\lam,a}^{}
:= -\frac{\rd}{\rd t}\Big|_{t=0} \e^{\i t h_\lam}\ia h_a^{}\ias
 \e^{-\i th_\lam}.
\ea
Moreover, its NESS expectation value is denoted by
\ba
\label{nessc-0}
J_{\lam,a}
:=\omega_{\lam,+}(\Phi_{\lam,a}).
\ea
\ed

Let us next turn to the structure of the NESS current. The following
proposition shows, on one hand, that the expectation value is independent
of the pure point component of the NESS density. On the other hand,
it implies that we can later proceed to its computation by exploiting the
known density of the XY NESS and the purely absolutely continuous nature
of the XY Hamiltonian in the construction of the wave operator.
The commutator of $A,B\in\mL(\mH)$ is denoted by $[A,B]:=AB-BA$. 
\bp[Energy current structure]
\label{prop:nessc}
For $a\in\{L,R\}$, we have
\ba
\label{nessc-1}
J_{\lam,a}
=\tr(w^\ast(h,h_\lam)\rho_{1,+}^{} w(h,h_\lam)\varphi_{\lam,a}^{}).
\ea
\ep
\bprf
Since $\varphi_{\lam,a}\in\mL^0(\fh)$ and using the form
\eqref{Skp} of the NESS density, we can write
\ba
\label{nessc-2}
J_{\lam,a}
&=\tr(\rho_{\lam,+}\varphi_{\lam,a}^{})\nonumber\\
&=\tr(w^\ast(h_0,h_\lambda )\rho w(h_0,h_\lambda)\varphi_{\lam,a}^{})
+\sum_{e\in\, \spec_\rpp(h_\lambda )}
\tr(1_e(h_\lambda)\rho 1_e(h_\lambda)\varphi_{\lam,a}^{}).
\ea
The independence of the current of the pure point component of the NESS density now follows as in Aschbacher {\it et al.} \cite{AJPP2} from
the observation that, since the one-particle energy current observable from \eqref{1eco-0}  has the form of a commutator, namely 
$\varphi_{\lam,a}^{}=-\i[h_\lam, \ia h_a^{}\ias]$, we have, 
for all $e\in\spec_\rpp(h_\lam)$, that
\ba
1_e(h_\lam)\varphi_{\lam,a}1_e(h_\lam)
=0.
\ea
Applying the chain rule $w(h_0,h_\lam)=w(h_0,h)w(h,h_\lam)$ 
to the wave operators in \eqref{nessc-2} (which
is applicable since the perturbations are trace class) and using
\eqref{rho1+-1}, we get
\ba
w^\ast(h_0,h_\lam)\rho w(h_0,h_\lam)
=w^\ast(h,h_\lam)\rho_{1,+}w(h,h_\lam).
\ea
This is the assertion.
\eprf
\br
Note that, due to $\varphi_{\lam,a}^{}
=-\i[h_\lam,\ia h_a^{}\ias]
=-\i\lam[v_a^{},\ia h_a^{}\ias]\in\mL^0(\fh)$,
we have
 $
\sum_{a\in\{L,R\}}\varphi_{\lam,a}^{}
=\i [h_\lam, \is h_\mS^{}\iss+\lam v]$, 
where $\is h_\mS^{}\iss+\lam v\in\mL^0(\fh)$. Hence, 
$\rd\Gamma(\is h_\mS^{}\iss+\lam v)\in\mO$, and 
\ba
\sum_{a\in\{L,R\}}\Phi_{\lam,a}^{}
=\frac{\rd}{\rd t}\Big|_{t=0} \tau_\lam^t
(\rd\Gamma(\is h_\mS^{}\iss+\lam v)).
\ea
Since Definition \ref{def:ness} implies that the NESS is invariant
under the corresponding $C^\ast$ dynamics, {\it i.e.} since
$\omega_{\lam,+}\circ\tau_\lam^t=\omega_{\lam,+}$ for all 
$\lam,t\in\R$, we get the first law of thermodynamics of the microscopic
regime,
\ba
\sum_{a\in\{L,R\}}J_{\lam,a}
=0.
\ea
Therefore, we can
restrict ourselves to study the objects from Definition 
\ref{def:current} for $a=L$, and, for this case, we drop the index $L$ in
the notation.
\er

We now arrive at the first of our main theorems. For any coupling strength and any size of the sample, it yields an explicit expression for the NESS energy current and, thus, for the microscopic entropy production (see, for example, Aschbacher {\it et al.} \cite{AJPP1}),
\ba
\Ep_\lam
:=-\sum_{a\in\{L,R\}} \beta_a J_{\lam,a}
=2\delta J_\lam.
\ea
Of course, for $\lam=0$, we have $\Ep_0=0$. If  $\lam\neq0$, we have the
following result.
\bt[Microscopic second law of thermodynamics]
\label{thm:nessc}
For $\lam\neq0$, the microscopic entropy production  is given by the absolutely convergent integral
\ba
\label{nessc-3}
\Ep_\lam
=\delta\lambda^4
\int_{-\pi}^\pi \,\,
\frac{\rd k}{2\pi}\,\, 
\frac{S(|\eps(k)|)}
{Q_\lam(|\eps(k)|)},
\ea
where the functions $S:[-1,1]\to\R$ and $Q_\lam: (-1,1)\to\R$ are 
defined by
\ba
\label{S}
S(e)
&:=e(1-e^2)^{\!1/2}[\planck_{\beta_L}(e)-\planck_{\beta_R}(e)],\\
\label{Q-0}
Q_\lam
&:=|(1-E^2)^{-2}[(1-\lam^2 E^2)^2-(1-\lam^2)^2 E^{2(n_\mS+1)}]|^2,
\ea
and $E(e):=e+\i(1-e^2)^{\!1/2}$.
Thus, if the system is truly out of equilibrium, {\it i.e.} if 
$\delta\neq 0$, the microscopic entropy production is strictly positive
and the energy current is flowing through the sample from the hotter to the colder reservoir. 
\et
\br
\label{rem:AP}
We can rewrite the microscopic entropy production in the form
\ba
\label{nessc-4}
\Ep_\lam
=\frac{\delta\lam^4}{2} \int_{-\pi}^\pi \,\,
\frac{\rd k}{2\pi}\,\, \frac{\eps(k)|\eps'(k)|}{Q_\lambda (\eps(k))}\,\,
\frac{\sh[\delta\eps(k)]}
{\ch^2[\tfrac\beta2\eps(k)]+\sh^2[\tfrac\delta2\eps(k)]},
\ea
where we used the convenient identity 
$2[(1+\e^x)^{-1}-(1+\e^y)^{-1}]=\sh[(y-x)/2]/(\ch^2[(x+y)/4]+\sh^2[(x-y)/4])$ for $x,y\in\R$. For $\lam=1$, we have $Q_1=1$ and, hence, we recover the expression found in Aschbacher and Pillet \cite{AP} (which, in addition, is also
independent of the sample size).
\er
\bprf
In order to analyze the NESS current, we start from  
\eqref{nessc-1} and determine its ingredients. For convenience, we will
work with the objects for $a=R$ in this proof. First, using 
\eqref{1eco-0}, we can write the one-particle energy current observable 
as
\ba
\label{1eco-1}
\varphi_{\lam,R}^{}
=\frac\lam2\, \Im[(\delta_n,\cdot\,)\,\delta_{n+2}].
\ea
Plugging \eqref{1eco-1} into \eqref{nessc-1}, we get
\ba
\label{nessc-4}
J_{\lam,R}
=\frac\lam2\, \Im[F_\lam(n, n+2)],
\ea
where, for all $\lam\in\R$, 
the function $F_\lam: \Z^2\to\C$ is defined by
\ba
\label{Fl-0}
F_\lam(x,y)
:=(w(h,h_\lam)\delta_x, \rho_{1,+}^{} w(h,h_\lam)\delta_y).
\ea
In order to compute this function and, in particular, the wave operator
appearing in it, we switch to the energy space of the XY
Hamiltonian ({\it i.e.} to the space diagonalizing $h$) 
given in Lemma \ref{lem:xy} of Appendix \ref{app:spec} by 
\ba
\wt\fh
=L^2([-1,1],\C^2;\rd e).
\ea
In this representation, the action of the wave operator on the completely localized wave functions $\delta_x\in\fh$ with $x\in\Z$  reads
\ba
\label{wave-3}
\wt w(h,h_\lam) \wt\delta_x
=\wt \delta_x
-\frac{\lam-1}{2}\sum_{i,j=1}^4 \Sigma_{\lambda, ij}^{-1}(\,\cdot-\i0)\,
\varrho_{\delta^1_j,\delta_x}(\,\cdot-\i0)\,\wt\delta^2_i,
\ea
where $\Sigma_\lambda(e-\i0)\in\C^{4\times 4}$ and 
$\varrho_{\delta^1_j,\delta_x}(e-\i0)\in\C$ are the boundary values of
the interaction matrix and of the XY resolvent amplitudes, respectively, 
and $\delta^1_i,\delta^2_i\in\fh$ are the coupling functions, see 
Proposition \ref{prop:wave-diag} of Appendix \ref{app:wave}.
Plugging \eqref{wave-3} into \eqref{Fl-0}, we get 
\ba
\label{mv-0}
F_\lam(x,y)
=\int_{-1}^1\rd e\,\,S^{(0)}_{x,y}(e)
+\sum_{i=1}^2 \left(\!\frac{1-\lam}{2}\!\right)^{\hspace{-0.8mm}i}
\int_{-1}^1\rd e\,\,S^{(i)}_{\lam,x,y}(e),
\ea
where, for all $\lam\in\R$, $x,y\in\Z$, and $i=1,2$, the functions 
$S^{(0)}_{x,y}, S^{(i)}_{\lam,x,y}:(-1,1)\to\C$ read
\ba
\label{S0-0}
S^{(0)}_{x,y}
&:={\big\langle\wt\delta_x, 
\wt\rho_{1,+}\wt\delta_y\big\rangle}_{\!2},\\
\label{S1-0}
S^{(1)}_{\lam,x,y}
&:={\big\langle \xi_x,\Sigma_\lam^{-1}(\,\cdot-\i 0)\eta_y\big\rangle}_{\!4}
+{\big\langle \Sigma_\lam^{-1}(\,\cdot-\i 0)\eta_x, \xi_y\big\rangle}_{\!4},\\
\label{S2-0}
S^{(2)}_{\lam,x,y}
&:={\big\langle\Sigma_\lam^{-1}(\,\cdot-\i 0)\eta_x,\Theta\Sigma_\lam^{-1}(\,\cdot-\i 0)\eta_y
\big\rangle}_{\!4}.
\ea
Here, for all $x\in\Z$, the vector-valued functions 
$\xi_x, \eta_x: (-1,1)\to\C^4$ and the matrix-valued function 
$\Theta:(-1,1)\to\C^{4\times 4}$ are defined, for $i,j=1,\ldots ,4$, by
\ba
\label{mv-1}
\xi_{x,i}
&:={\big\langle\wt\delta^2_i,
\wt\rho_{1,+}^{}\wt\delta_x\big\rangle}_{\!2},\\
\label{mv-2}
\eta_{x,i}
&:=\varrho_{\delta^1_i,\delta_x}(\,\cdot-\i 0),\\
\label{mv-3}
\Theta_{ij}
&:={\big\langle\wt\delta^2_i,\wt\rho_{1,+}^{}\wt\delta^2_j\big
\rangle}_{\!2},
\ea
where ${\langle\,\cdot\,,\cdot\,\rangle}_d$ stands for the Euclidean scalar product in $\C^d$. 
We next specialize to the case at hand, namely to $x=n$ and 
$y=n+2$. For this case, the ingredients of 
\eqref{S0-0}--\eqref{S2-0} are computed in Lemma \ref{lem:ingr} of
Appendix \ref{app:wave}. Plugging these expressions into 
\eqref{mv-0}, we get
\ba
\label{RatFct}
F_\lam(n,n+2)
= F_1(n,n+2)
+\int_{-1}^1\rd e\,\,
\frac{\sum_{i=0}^8 p_i(e)\lambda ^i}
{Q_\lam(e)},
\ea
where the function $Q_\lam: (-1,1)\to\R$, defined by
$Q_\lam:=|\hspace{-0.4mm}\det(\Sigma_\lam(\,\cdot-\i0)|^2$, has the expansion
$Q_\lam=\sum_{i=0}^4q_{2i}\,\lam^{2i}$, and the coefficient functions $p_i,q_i:(-1,1)\to\C$  are given in Lemma 
\ref{lem:coeff} of Appendix \ref{app:wave}. Subtracting $F_0(n,n+2)=0$ from \eqref{RatFct} (where the latter follows from Lemma \ref{lem:coeff}, see also \eqref{Fl-0}), we can write
\ba
\label{C-3}
F_\lam(n,n+2)
=\int_{-1}^1\rd e\,\,  \frac{\sum_{i=0}^3 p_{2i+1}(e)\lam^{2i+1}
}{Q_\lam(e)},
\ea
where we used  that $q_0p_{2i}-p_0q_{2i}=0$ for $i=1,\ldots ,4$, see
Lemma \ref{lem:coeff} of Appendix \ref{app:wave}. Transforming the coordinates as $e=\eps(k)$ for $k\in[0,\pi]$ and using that 
$\Im[p_{2i+1}(\eps(k))]=0$ for $i=0,2,3$ and 
$\Im[p_{3}(\eps(k))]=-\eps(k)
[\planck_{\beta_L}(\eps(k))-\planck_{\beta_R}(\eps(k))]/\pi$ from 
Lemma \ref{lem:coeff}, we get \eqref{nessc-3}. 
Finally, due to Lemma \ref{lem:roots} in Appendix \ref{app:wave},
the integral in \eqref{nessc-3} is absolutely convergent, and since
the numerator and the denominator are even functions in $e$, we arrive 
at the assertion.
\eprf
\section{Leading order microscopic regime}
\label{sec:macro} 

\begin{figure}
\centering
\includegraphics[width=7cm,height=5cm]{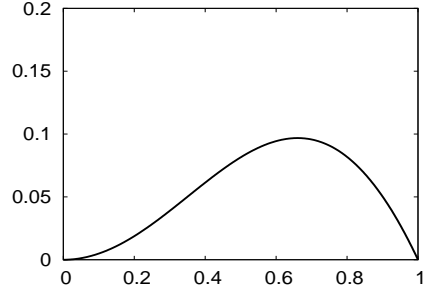}
\caption{The function $S_0$ on $[0,1]$ for $\beta_L=1$ and 
$\beta_R=2$.}
\label{fig:gnu-current-function}
\end{figure}

In this section, we determine the leading order contribution to 
the microscopic entropy production  from Theorem \ref{thm:nessc} for small bond coupling $\lam$. It has the following form.
\bt[Leading order contribution]
\label{thm:2nd-order}
For $\lam\to0$, the microscopic entropy production has the expansion
\ba
\Ep_\lam
=\Ep\,\lam^2+\mO(\lam^4),
\ea
where the second order contribution has the form
\ba
\label{Ja2}
\Ep
:=\frac{2\delta}{n_\mS+1}
\sum_{i=1}^{n_\mS} 
S_0(\eps(k_i)),
\ea
and the function $S_0:[-1,1]\to\R$ (see Figure \ref{fig:gnu-current-function}) and the momenta $k_i$ for 
$i=1,\ldots ,n_\mS$ read
\ba
\label{vtheta}
S_0(e)
&:=e(1-e^2)^{\!3/2}[\planck_{\beta_L}(e)-\planck_{\beta_R}(e)],\\
k_i
&:=\frac{i\pi}{n_\mS+1}.
\ea
\et
\br
Note that $\eps(k_{n+1+i})=-\eps(k_{n+1-i})$ for $i=1,\ldots ,n$
and that $S_0$ is an even function. Hence, for $n>0$, we can further simplify \eqref{Ja2} as $\Ep=2\delta/(n+1)\sum_{i=1}^{n} S_0(\eps(k_i))$.
\er
\br
It follows from $\eps(k_{n+1})=0$ that, for $n=0$, we have $\Ep=0$.
Hence, in this case, the entropy production from Theorem \ref{thm:nessc} is carried by higher orders than the second one. On the other hand, since 
$S_0\circ\eps\in C([-\pi,\pi];\tfrac{\rd k}{2\pi})$ is an even function, 
we get with the identity from Remark \ref{rem:AP} that
\ba
\lim_{n\to\infty} \Ep
=\delta\int_{-\pi}^\pi \,\,
\frac{\rd k}{2\pi}\,\, \eps(k) |\eps'(k)|^3\,\,
\frac{\sh[\delta\eps(k)]}
{\ch^2[\tfrac\beta2\eps(k)]+\sh^2[\tfrac\delta2\eps(k)]}.
\ea
\er

Let us now turn to the proof of Theorem \ref{thm:2nd-order}.

\bprf
In order to extract the second order contribution to the NESS
current, we determine the limit $\lam \to0$ of $J_\lam/\lam^2$ with
the help of a Sokhotski-Plemelj type argument. For this
purpose, we rewrite the quotient in \eqref{nessc-3} as 
$S/Q_\lam=N/D_\lam$, where 
$N,D_\lam:\R\to\R$ read 
\ba
\label{rN}
N
&:=4(1-\eps^2)^2\,S(|\eps|),\\
\label{rD}
D_\lam
&:=L_\lam+\lam^4 R_\lam,
\ea
and the functions $L_\lam,R_\lam :\R\to\R$ are given by
\ba
\label{rL}
L_\lam 
&:= \sum_{i\in\{0,2\}} d_i\lam^i+d_{4'}\lam^4,\\
\label{rR}
R_\lam 
&:= \sum_{i\in\{4,6,8\}} d_i\lam^{4-i}-d_{4'}.
\ea
Moreover, the coefficients $d_i:\R\to\R$ for $i=0,2,4,4',6,8$ have the form
\ba
d_0
&=\sigma_{n_0}^2,\\
d_2
&=-4\sigma_{n_0}\sigma_{n_1}\eps,\\
d_4
&=2\sigma_{n_0}\sigma_{n_2}\eps_2+4\sigma^2_{n_1},\\
d_{4'}
&:=4\sigma^2_{n_1}\eps^2,\\
d_6
&=-4\sigma_{n_1}\sigma_{n_2}\eps,\\
d_8
&=\sigma^2_{n_2},
\ea
where, for all $\alpha\in\R$, we used the notation 
$\sigma_\alpha(k):=\sin(\alpha k)$, $\sigma:=\sigma_1$,  and 
$\eps_\alpha(k):=\eps(\alpha k)$
for all $k\in\R$, and we set $n_0:=n_\mS+1$, $n_1:=n_\mS$, and 
$n_2:=n_\mS-1$. Moreover, from now on, if not stated otherwise, we always assume that $|\lam|>0$.
In order to apply the Sokhotski-Plemelj  argument, we analyze the neighborhoods of the roots of $d_0$, located at
\ba
k_x
:=\frac{x\pi}{n_0},
\ea
where $x\in M:=\{x\in\Z\,|\, |x|\le n_\mS+1\}$. The neighborhoods of these roots are denoted by $K_x:=(k_x-\kp_x,k_x+\kp_x)\cap 
(-\pi,\pi)$ and their size $\kp_x$, satisfying $0<\kp_x\le k_1/2$ for all $x\in M$, will be suitably chosen below. Moreover, for all $x\in M$, 
we define the integrals 
\ba
\label{I}
I_{\lam,x}
:=\lam^2\int_{K_x}\frac{\rd k}{2\pi}\, \frac{N(k)}{D_\lam (k)}.
\ea
Then, we can make the decomposition
\ba
\label{curr-1}
\frac{2J_\lam}{\lam^2}
=\sum_{x\in M_0} I_{\lam,x}
+\sum_{x\in M\setminus M_0} I_{\lam,x}
+\lam^2\int_{K^c}\frac{\rd k}{2\pi}\, \frac{N(k)}{D_\lam(k)},
\ea
where we set $M_0:=\{0,\pm n_0\}$ and  
$K^c:= (-\pi,\pi)\setminus(\cup_{x\in M} K_x)$.
In the following, we will successively study all the contributions 
of the different integration domains in the decomposition \eqref{curr-1}. 
The coupling strength will always be assumed sufficiently small without
necessarily specifying its size in each estimate. Moreover,
if nothing else is indicated, the estimates are supposed to hold for 
all momenta.
Finally, the positive constant $C$ can take different values 
at each place it appears.

{\it Case 1:}\, $K_x$ for $x\in M_0$\\
Let us set 
$\kp_x:=\kp_0
:=k_1/2$ 
for all $x\in M_0$ and let us
rewrite \eqref{rN} by using the identity from Remark 
\ref{rem:AP}. 
Then, after an eventual shift of \eqref{I} to the origin, 
we have $|N(k_x+k)|\le Ck^2$ and  $d_0(k_x+k)\ge Ck^2$ 
for $|k|<\kp_0$. Since $\sum_{i=1}^4|d_{2i}(k_x+k)|\le Ck^2$, 
we get $D_\lam(k_x+k)\ge Ck^2$ for $|k|<\kp_0$. Hence, for $x\in M_0$ 
and $\lam\to 0$, we find 
\ba
\label{Ic1}
I_{\lam,x}
=\mO(\lam^2).
\ea

{\it Case 2:}\, $K_x$ for $x\in M\setminus M_0$\\
In order to determine some size $\kp_x$ for the neighborhood $K_x$, 
we estimate \eqref{rR} from below as follows. First, 
we define the function $d_{4''}:\R\to\R$ by 
$d_{4''}:=4\sigma^2_{n_1}\sigma^2$ which allows us to write 
\eqref{rR}, shifted to the origin, in the form 
\ba
R_\lam(k_x+k)
&=d_{4''}(k_x)\\
\label{R-1}
&+\,d_{4''}(k_x+k)-d_{4''}(k_x)\\
\label{R-2}
&+\,d_4(k_x+k)-d_{4'}(k_x+k)-d_{4''}(k_x+k)\\
\label{R-3}
&+\,\lam^2 d_6(k_x+k)\\
\label{R-4}
&+\,\lam^4 d_8(k_x+k).
\ea
Using $|\eqref{R-1}|, |\eqref{R-2}|\le C |k|$ and 
$|\eqref{R-3}|, |\eqref{R-4}| \le C$, we have
$R_\lam(k_x+k)\ge 4 \sigma^4(k_x)-C(|k|+\lam^2)$.
Moreover, we note that $\sigma^4(k_x)\ge \sigma^4(k_1)=:\vartheta>0$ for all $x\in M\setminus M_0$. Hence, there exists a $\kp>0$ which we can choose as $\kp:=\vartheta/(3 n_0)\le 1/6$, s.t., for $|k|<\kp$, we have
\ba
\label{rR-est}
R_\lam(k_x+k)
\ge \vartheta.
\ea
Therefore, we set $\kp_x:=\kp$ for all $x\in M\setminus M_0$.   In order
to study \eqref{I} in the neighborhood $K_x=(k_x-\kp,k_x+\kp)$,
we make the decomposition $I_{\lam,x}=A_{\lam,x}+N(k_x) B_{\lam,x}/n_0$, 
where 
\ba
\label{A}
A_{\lam,x}
&:= \lam^2 \int_{K_x}\frac{\rd k}{2\pi}\, \frac{N(k)-N(k_x)}
{D_\lam (k)},\\
\label{B}
B_{\lam,x}
&:= n_0\lam^2 \int_{K_x}\frac{\rd k}{2\pi}\,\frac{1}{D_\lam (k)}.
\ea
Let us first analyze \eqref{A}. To this end, we further decompose it as $A_{\lam,x}=A_{\lam,x}^{(1)}+A_{\lam,x}^{(2)}$, where
\ba
\label{A-1}
A^{(1)}_{\lam,x}
&:=\lam^2 \int_0^\kp\frac{\rd k}{2\pi}\,
\frac{N_x^+(k)L_\lam(k_x-k)+N_x^-(k)L_\lam(k_x+k)}
{D_\lam(k_x+k)D_\lam(k_x-k)},\\
\label{A-2}
A^{(2)}_{\lam,x}
&:=\lam^6 \int_0^\kp\frac{\rd k}{2\pi}\,
\frac{N_x^+(k)R_\lam(k_x-k)+N_x^-(k)R_\lam(k_x+k)}
{D_\lam(k_x+k)D_\lam(k_x-k)},
\ea
and the functions $N_x^\pm:\R\to\R$ are defined by 
$N_x^\pm(k):=N(k_x\pm k)-N(k_x)$. 
Next, let us decompose \eqref{A-1} as 
$A^{(1)}_{\lam,x}=A^{(1,1)}_{\lam,x}+A^{(1,2)}_{\lam,x}$, where 
\ba
\label{A-11}
A^{(1,1)}_{\lam,x}
&:=\lam^2 \int_0^\kp\frac{\rd k}{2\pi}\,
\frac{N_x^+(k)d_0(k_x-k)+N_x^-(k)d_0(k_x+k)}
{D_\lam(k_x+k)D_\lam(k_x-k)},\\
\label{Am-12}
A^{(1,2)}_{\lam,x}
&:=\lam^4 \int_0^\kp\frac{\rd k}{2\pi}\,
\frac{N_x^+(k)(d_2+\lam^2d_{4'})(k_x-k)
+N_x^-(k)(d_2+\lam^2d_{4'})(k_x+k)}{D_\lam(k_x+k)D_\lam(k_x-k)}.
\ea
In order to bound \eqref{A-11},
we use $d_0(k_x\pm k)=d_0(k)$ for all $x\in M$, 
 expand $N\in C^\infty(\R\setminus \pi\Z)$ around $k_x$ up to second
order in $\pm k$, and apply $D_\lam(k_x\pm k)\ge L_\lam(k_x\pm k)+\vartheta\lam^4>0$ for all $|k|< \kp$ 
which follows from \eqref{rR-est} and the fact that
$L_\lam=\ell_\lam^2\ge 0$, where 
$\ell_\lam:\R\to\R$ is 
\ba
\label{ell}
\ell_\lam
:=\sigma_{n_0}-2\lam^2\sigma_{n_1}\eps. 
\ea
Hence, we get
\ba
\label{A11-e1}
|A^{(1,1)}_{\lam,x}|
\le C \lam^2 \int_0^\kp\frac{\rd k}{2\pi}\,
\frac{k^2d_0(k)}
{([\ell_\lam(k_x+k)]^2+\vartheta\lam^4)([\ell_\lam(k_x-k)]^2+\vartheta\lam^4)}.
\ea
Since $d_0=\sigma^2_{n_0}$ and $\ell_\lam(k_x\pm k)=
\sigma_{n_0}(k)\pm 2(-1)^{x+1}\lam^2\sigma(n_1 [k_x\pm k])
\eps(k_x\pm k)$, we make the coordinate transformation 
$k=\arcsin(\lam^2 p)/n_0$ for 
$0\le p\le a_\lam:=\sigma(n_0\kp)/\lam^2$, where $n_0\kp
<\pi/2$. Hence, we get
\ba
|\mbox{r.h.s. } \eqref{A11-e1}|
\label{A11-e2}
\le C \int_0^{a_\lam}\frac{\rd p}{2\pi}\,
\frac{p^2[\arcsin(\lam^2 p)]^2}
{([p+\ell_{\lam,x}^+(p)]^2+\vartheta)([p+\ell_{\lam,x}^-(p)]^2+\vartheta)},
\ea
where, for $\Lambda_\lam:=(-1/\lam^2,1/\lam^2)$, 
the functions $\ell_{\lam,x}^\pm:\Lambda_\lam\to \R$ are defined  by 
\ba
\ell_{\lam,x}^\pm(p)
&:=[\ell_1-\ell_0](k_x\pm\arcsin(\lam^2 p)/n_0)\nonumber\\
&=\pm 2(-1)^{x+1}\sigma(n_1 [k_x\pm \arcsin(\lam^2 p)/n_0])
\eps(k_x\pm \arcsin(\lam^2 p)/n_0).
\ea
Since $|\ell_{\lam,x}^\pm(p)|\le 2$ for all $p\in \Lambda_\lam$, we find
\ba
|\mbox{r.h.s. } \eqref{A11-e2}|
\label{A11-e3}
\le C \lam^4 \left(\int_0^2\frac{\rd p}{2\pi}\,\frac{p^4}{\vartheta^2}
+\int_2^{a_\lam}\frac{\rd p}{2\pi}\,\frac{p^4}{([p-2]^2
+\vartheta)^2}\right).
\ea
The integrand of the second
integral on the r.h.s. of \eqref{A11-e3} is bounded and, hence, due to
$a_\lam=\sigma(\vartheta/3)/\lam^2$, we find 
$A^{(1,1)}_{\lam,x}=\mO(\lam^2)$.
We next turn to the estimate of \eqref{Am-12}. For this purpose, we make
the decomposition 
$A^{(1,2)}_{\lam,x}=A^{(1,2,1)}_{\lam,x}+A^{(1,2,2)}_{\lam,x}$, where 
\ba
\label{A121}
A^{(1,2,1)}_{\lam,x}
&:= \lam^4 \int_0^\kp\frac{\rd k}{2\pi}\,
\frac{N_x^+(k)d_2(k_x-k)
+N_x^-(k)d_2(k_x+k)}{D_\lam(k_x+k)D_\lam(k_x-k)},\\
\label{A122}
A^{(1,2,2)}_{\lam,x}
&:= \lam^6 \int_0^\kp\frac{\rd k}{2\pi}\,
\frac{N_x^+(k)d_{4'}(k_x-k)
+N_x^-(k)d_{4'}(k_x+k)}{D_\lam(k_x+k)D_\lam(k_x-k)}.
\ea
In order to bound \eqref{A121}, we first bound the numerator in 
\eqref{A121} by using 
$|N_x^\pm(k)|\le 
Ck$ and 
$|d_2(k_x\pm k)|\le 4 \sigma(n_0 k)$ for all $k\in [0,\kp]$, and 
we treat the denominator in \eqref{A121} as in \eqref{A11-e1}. Then, proceeding as in \eqref{A11-e2}, we can write
\ba
\label{A121-e1}
|A_x^{(1,2,1)}(\lam)|
\le C \int_0^{a_\lam}\frac{\rd p}{2\pi}\, \frac{p \arcsin(\lam^2 p)}
{([p+\ell_{\lam,x}^+(p)]^2+\vartheta)
([p+\ell_{\lam,x}^-(p)]^2+\vartheta)}.
\ea
Moreover, analogously to \eqref{A11-e3}, we get 
\ba
|\mbox{r.h.s. } \eqref{A121-e1}|
\label{A121-e2}
\le C \lam^2 \left(\int_0^2\frac{\rd p}{2\pi}\,
\frac{p^2}{\vartheta^2}
+\int_2^{a_\lam}\frac{\rd p}{2\pi}\,\frac{p^2}
{([p-2]^2+\vartheta)^2}\right).
\ea
Extending the integration domain of the second integral on the r.h.s. of \eqref{A121-e2} to infinity, we get
$A^{(1,2,1)}_{\lam,x}=\mO(\lam^2)$.
In order to bound \eqref{A122}, using $|d_{4'}(k_x\pm k)|\le C$  and again $|N_x^\pm(k)|\le 
Ck$ for all $k\in [0,\kp]$,
we can proceed as above and get
\ba
\label{A122-e1}
|A^{(1,2,2)}_{\lam,x}|
\le C \lam^2 \left(\int_0^2\frac{\rd p}{2\pi}\,\frac{p}{\vartheta^2}
+\int_2^{a_\lam}\frac{\rd p}{2\pi}\,\frac{p}
{([p-2]^2+\vartheta)^2}\right),
\ea
which again implies that $A^{(1,2,2)}_{\lam,x}=\mO(\lam^2)$. Finally,
in order to bound \eqref{A-2},
we note that  $|R_\lam(k_x\pm k)|\le C$, and estimating \eqref{A-2} as \eqref{A122}, we get $A^{(2)}_{\lam,x}=\mO(\lam^2)$. 
Taking all of the foregoing estimates together finally implies
that the term \eqref{A} does not contribute anything to the second order
of the current, {\it i.e.}, for $\lam\to 0$, we have 
$A_{\lam,x} =\mO(\lam^2)$ for all $x\in M\setminus M_0$.
We next turn to the study of \eqref{B}. 
For this purpose, we rewrite \eqref{B} using the coordinate transformation introduced before \eqref{A11-e2} which leads to
\ba
\label{B-1}
B_{\lam,x}
=\int_{-a_\lam}^{a_\lam}\frac{\rd p}{2\pi}\,\, Y_{\lam,x}(p),
\ea
where  the function $Y_{\lam,x}: \Lambda_\lam\to\R$ is defined by
\ba
\label{rQ}
Y_{\lam,x}(p)
:=\frac{1}{(1-\lam^4p^2)^{1/2}}\,
\frac{1}{[p+\ell_{\lam,x}(p)]^2+R_{\lam,x}(p)},
\ea
and we set $\ell_{\lam,x}:=\ell_{\lam,x}^+$, and 
the function $R_{\lam,x}: \Lambda_\lam\to\R$ is given by 
$R_{\lam,x}(p):=R_\lam(k_x+\arcsin(\lam^2 p)/n_0)$.
Let us first write 
$B_{\lam,x}=B_{0,x}+\big[B_{\lam,x}-B_{0,x}\big]$, where 
\ba
\label{B0}
B_{0,x}
:=\int_{-\infty}^\infty\frac{\rd p}{2\pi}\,\, Y_{0,x}(p),
\ea
the function $Y_{0,x}:\R\to\R$  is defined by
\ba
\label{rQ0}
Y_{0,x}(p)
:=\frac{1}{[p+\ell_{0,x}]^2+R_{0,x}},
\ea
and the constants are given, for all $x\in M\setminus M_0$, by
$\ell_{0,x}:=\sigma_2(k_x)$ and $R_{0,x}:=4\sigma^4(k_x)>0$.
Let us next decompose the difference as
$B_{\lam,x}-B_{0,x}
=B^{(1)}_{\lam,x}-B^{(2)}_{\lam,x}$, where 
\ba
\label{B1}
B^{(1)}_{\lam,x}
&:=\int_{-a_\lam}^{a_\lam}\frac{\rd p}{2\pi}\, 
\big[Y_{\lam,x}(p)-Y_{0,x}(p)\big],\\
\label{B2}
B^{(2)}_{\lam,x}
&:=\int_{|p|\ge a_\lam}\frac{\rd p}{2\pi}\, Y_{0,x}(p).
\ea
Furthermore, we make the  decomposition
$B^{(1)}_{\lam,x}=B^{(1,1)}_{\lam,x}+B^{(1,2)}_{\lam,x}$, where 
\ba
\label{B11}
B^{(1,1)}_{\lam,x}
&:= \lam^4\int_{-a_\lam}^{a_\lam}\frac{\rd p}{2\pi}\, 
\frac{p^2}{(1-\lam^4p^2)^{1/2}(1+(1-\lam^4p^2)^{1/2})}
\frac{1}{[p+\ell_{\lam,x}(p)]^2+R_{\lam,x}(p)},\\
\label{B12}
B^{(1,2)}_{\lam,x}
&:=\int_{-a_\lam}^{a_\lam}\frac{\rd p}{2\pi}\, 
\left(\frac{1}{[p+\ell_{\lam,x}(p)]^2+R_{\lam,x}(p)}
-\frac{1}{[p+\ell_{0,x}]^2+R_{0,x}}\right).
\ea
In order to bound \eqref{B11}, we use $(1-\lam^4p^2)^{1/2}\ge \eps(\vartheta/3)$ for all $|p|\le a_\lam$,
$|\ell_{\lam,x}(p)|\le 2$, and \eqref{rR-est} which yields
\ba
|B^{(1,1)}_{\lam,x}|
\le C \lam^4\left(\int_0^2\frac{\rd p}{2\pi}\,
\frac{p^2}{\vartheta}+\int_2^{a_\lam}\frac{\rd p}{2\pi}\,
\frac{p^2}{(p-2)^2+\vartheta}\right).
\ea
Due to the boundedness of the integrand of the second integral, we have $B^{(1,1)}_{\lam,x}=\mO(\lam^2)$. In order to estimate \eqref{B12}, we make the decomposition $B^{(1,2)}_{\lam,x}=B^{(1,2,1)}_{\lam,x}+B^{(1,2,2)}_{\lam,x}$, where
\ba
\label{B121}
B^{(1,2,1)}_{\lam,x}
&:=\int_{-a_\lam}^{a_\lam}\frac{\rd p}{2\pi}\, 
\frac{[p+\ell_{0,x}]^2-[p+\ell_{\lam,x}(p)]^2}
{([p+\ell_{0,x}]^2+R_{0,x})([p+\ell_{\lam,x}(p)]^2
+R_{\lam,x}(p))},
\\
\label{B122}
B^{(1,2,2)}_{\lam,x}
&:=\int_{-a_\lam}^{a_\lam}\frac{\rd p}{2\pi}\, 
\frac{R_{0,x}-R_{\lam,x}(p)}
{([p+\ell_{0,x}]^2+R_{0,x})([p+\ell_{\lam,x}(p)]^2
+R_{\lam,x}(p))}.
\ea
In order to bound \eqref{B121}, we use
$|[p+\ell_{0,x}]^2-[p+\ell_{\lam,x}(p)]^2|
\le C(1+|p|) |\ell_{0,x}-\ell_{\lam,x}(p)|$
and $|\ell_{0,x}-\ell_{\lam,x}(p)|\le C\lam^2 |p|$ 
for the numerator and, for the denominator,
$|\ell_{0,x}|,|\ell_{\lam,x}(p)|\le 2$  and \eqref{rR-est} which implies
\ba
|B^{(1,2,1)}_{\lam,x}|
\le C \lam^2\left(\int_0^2\frac{\rd p}{2\pi}\,\frac{p}{\vartheta^2}
+\int_2^{a_\lam}\frac{\rd p}{2\pi}\,
\frac{p^2}{([p-2]^2+\vartheta)^2}\right).
\ea
Extending the second integral to infinity, we get
$B^{(1,2,1)}_{\lam,x}=\mO(\lam^2)$.
In order to bound \eqref{B122}, we use 
$|R_{0,x}-R_{\lam,x}(p)|\le C(1+|p|)\lam^2$ and estimate the denominator as above yielding
\ba
|B^{(1,2,2)}_{\lam,x}|
\le C\lam^2\left(\int_0^2 \frac{\rd p}{2\pi}\, \frac{1}{\vartheta^2}
+\int_2^{a_\lam}\frac{\rd p}{2\pi}\,
\frac{p}{([p-2]^2+\vartheta)^2}\right).
\ea
Extending the second integral to infinity, we again get 
$B^{(1,2,2)}_{\lam,x}=\mO(\lam^2)$. 
We next turn to the estimate of \eqref{B2}.
Again, from $|\ell_{0,x}|\le 2$  and $R_{0,x}\ge\vartheta$, we have
\ba
|B^{(2)}_{\lam,x}|
\le C \int_{a_\lam}^\infty \frac{\rd p}{2\pi}\, \frac{1}{[p-2]^2+\vartheta},
\ea
and a coordinate transformation fixing the lower limit of the integral
 leads to $B^{(2)}_{\lam,x}=\mO(\lam^2)$.
Collecting the estimates for \eqref{A} and \eqref{B}, we get that, 
for $x\in M\setminus M_0$ and $\lam\to0$,
\ba
\label{Ic2}
I_{\lam,x}
=\frac{N(k_x)}{n_0}\int_{-\infty}^\infty\frac{\rd p}{2\pi}\, Y_{0,x}(p)
+\mO(\lam^2).
\ea
It remains to study the last term in \eqref{curr-1}.

{\it Case 3:}\, $K^c$\\
Rewriting the numerator of the integrand with the help of the identity
from Remark \ref{rem:AP}, we immediately get
$|N(k)|\le C$.
Moreover, the zeroth order contribution of the denominator is bounded from below by $d_0(k)\ge d_0(\kp)>0$ for all $k\in K^c$. Since we have 
$|D_\lam(k)-d_0(k)|\le C\lam^2$, we get $D_\lam(k)\ge d_0(\kp)/2$ for all
$k\in K^c$. Hence, for $\lam\to 0$,  we find 
\ba
\label{Ic3}
\int_{K^c}
\frac{\rd k}{2\pi}\, \frac{N(k)}{D_\lam(k)}
=\mO(1).
\ea

We can now extract the nontrivial second order contribution to the NESS current from \eqref{curr-1}.
Using \eqref{Ic1}, \eqref{Ic2}, and \eqref{Ic3}, 
it follows from \eqref{curr-1} that
\ba
\label{lim-1}
\lim_{\lam\to 0}
\frac{J_\lam}{\lam^2}
&=\frac12\sum_{x\in M\setminus M_0}\frac{N(k_x)}{n_0}
 \int_{-\infty}^\infty\frac{\rd p}{2\pi}\, 
Y_{0,x}(p)\nonumber\\
&=\frac{1}{4n_0}\sum_{x\in M\setminus M_0} 
\frac{N(k_x)}{R_{0,x}^{1/2}}.
\ea 
This concludes the proof of the theorem.
\eprf
\section{Van Hove regime}
\label{sec:vanHove}

We start this section by introducing what we call the product setting. In
this setting, the sample algebra is split off from the total algebra, and
we can conveniently focus on the thermodynamics of the sample system. It
is defined as follows.

\bd[Product setting]
\label{def:ps}
The ingredients for this setting, partially labelled by a tilde, are specified as follows.
\begin{enumerate}
\item[\textnormal{(a)}] \textnormal{Observable algebra}\\
The observable algebras of the sample and the reservoir are defined by
\ba
\mO_\mS
&:=\fA(\fh_\mS),\\
\mO_\mR
&:= \fA(\fh_\mR),
\ea
and the total observable algebra is defined to be their tensor product,
\ba
\tilde\mO
:=\mO_\mS\otimes\mO_\mR.
\ea

\item[\textnormal{(b)}] \textnormal{Dynamics}\\
The Hamiltonians of the sample, the reservoir, the decoupled, and the
coupled system are specified by
\ba
\label{Hs1}
H_\mS
&:= \dG(h_\mS),\\
H_\mR
&:= \dG(h_\mR),\\
\tilde H_0
&:= H_\mS\otimes 1 +1 \otimes H_\mR,\\
\tilde H_{\lam,a}
&:=\tilde H_0+\lam \tilde V_a,\\
\tilde H_\lam
&:=\tilde H_0+\lam \tilde V,
\ea
where the couplings $\tilde V_a, \tilde V\in\tilde\mO$ are given by
\ba
\label{V1a}
\tilde V_a
&:= -\i \Gamma(-1 )\otimes1 \,\Im[a^\ast(\iss\delta_{\mS,a})\otimes 
a(\irs\delta_{\mR,a})],\\
\label{V1}
\tilde V
&:=\sum_{a\in\{L,R\}} \tilde V_a.
\ea
Correspondingly, the dynamics $\tau_\mS^t\in\Aut(\mO_\mS)$,
$\tau_\mR^t\in\Aut(\mO_\mR)$, and 
$\tilde\tau_{\lam,a}^t, \tilde\tau_\lam^t\in\Aut(\tilde\mO)$ are given, for all $A$ in $\mO_\mS$, $\mO_\mR$, and $\tilde\mO$, respectively, by
\ba
\label{S-1}
\tau_\mS^t(A)
&:= \e^{\i t H_\mS} A\, \e^{-\i t H_\mS},\\
\label{Res-1}
\tau_\mR^t(A)
&:= \e^{\i t H_\mR} A\, \e^{-\i t H_\mR},\\
\label{lam-a}
\tilde\tau_{\lam,a}^t(A)
&:= \e^{\i t \tilde H_{\lam,a}} A\, \e^{-\i t \tilde H_{\lam,a}},\\
\label{lam-1}
\tilde\tau_\lam^t(A)
&:= \e^{\i t \tilde H_\lam} A\, \e^{-\i t \tilde H_\lam}.
\ea

\item[\textnormal{(c)}]  \textnormal{Initial state}\\
The initial state of the sample $\omega_\mS\in\mE(\mO_\mS)$
and of the reservoir $\omega_\mR\in\mE(\mO_\mR)$ are defined 
to be quasifree with the densities $\rho_\mS\in\mL(\fh_\mS)$ and 
$\rho_\mR\in\mL(\fh_\mR)$, respectively.
The total initial state $\tilde \omega\in\mE(\tilde\mO)$ is defined by
\ba
\tilde \omega
:= \omega_\mS\otimes\omega_\mR.
\ea
\end{enumerate}
\ed

In order to show the equivalence of this product setting and the 
quasifree setting from Definition \ref{def:qfs}, we make use of the following lemma. We denote by $\mU(\mH)$ the unitary operators on the Hilbert space $\mH$.
\bl[Exponential law for fermions]
\label{lem:exp}
For $i=1,2$, let $\fF(\fh_i)$ be the fermionic Fock spaces over the Hilbert spaces $\fh_i$ having vacua
$\Omega_i$, creation and annihilation operators 
$a_i^\ast, a_i:\fh_i\to\mL(\fF(\fh_i))$, and  second quantizations
$\Gamma_i: \mU(\fh_i)\to \mL(\fF(\fh_i))$. 
Then, there exists a unique 
$U\in\mU(\fF(\fh_1\oplus\fh_2),\fF(\fh_1)\otimes\fF(\fh_2))$ s.t. 
\ba
U\Omega
&=\Omega_1\otimes\Omega_2,\\
U a(f_1\oplus f_2) U^\ast
&= a_1(f_1)\otimes 1 _2 + \Gamma_1(-1 _1)\otimes a_2(f_2),\\
U\Gamma(U_1\oplus U_2)U^\ast
&= \Gamma_1(U_1)\otimes \Gamma_2(U_2),
\ea
where $\Omega$, $a^\ast$, $a$, and $\Gamma$ are the corresponding objects 
for $\fF(\fh_1\oplus\fh_2)$.
\el
\bprf
See, for example,  Alicki and Fannes \cite{AF}.
\eprf

The two settings are then equivalent in the following sense.
\bl[Product setting isomorphism]
\label{lem:p-i}
Let $\Phi: \mL(\fF(\fh_\mS\oplus\fh_\mR))\to \mL(\fF(\fh_\mS)\otimes
\fF(\fh_\mR))$ be defined by $\Phi(A):=U A U^\ast$, 
where 
$U\in\mU(\fF(\fh_\mS\oplus\fh_\mR),\fF(\fh_\mS)\otimes\fF(\fh_\mR))$ is the unitary from Lemma \ref{lem:exp} corresponding to the decomposition
$\fh\simeq\fh_\mS\oplus\fh_\mR$. Then, $\Phi$ is a $C^\ast$ algebra ${}^\ast$-isomorphism. Moreover, the following assertions hold.
\begin{enumerate}
\item[(a)] $\Phi: \mO\to\tilde\mO$ is a $C^\ast$ algebra 
${}^\ast$-isomorphism.
\item[(b)] $\tilde\tau_\lam^t=\Phi\circ\tau_\lam^t\circ\Phi^{-1}$ 
for all $\lam,t\in\R$
\item[(c)] $\tilde \omega=\omega\circ\Phi^{-1}$
\end{enumerate}
\el
\bprf
Note that for our sample of finite size, $n_\mS<\infty$, we have 
\ba
\Gamma(-1)
=\prod_{x\in\Z_\mS}(1 -2a^\ast(\iss\delta_x)a(\iss\delta_x))
\in \mO_\mS. 
\ea
Moreover, the couplings are related by
$\Phi(V_\alpha)=\tilde V_\alpha$.
The proof is then analogous to the one of Aschbacher {\it et al.} 
\cite{AJPP1}, see there for details. 
\eprf

We next specify the van Hove weak coupling regime (see also Aschbacher {\it et al.} \cite{AJPP1} for example). For this purpose, we make use of the weak coupling theory developed by Davies \cite{Davies1,Davies2} and 
summarized for our needs in Appendix \ref{app:davies}.
\bd[Van Hove regime]
\label{def:macro}
Let the operator $P_\mS: \tilde\mO\to \mO_\mS$ be defined, for all $A\in \mO_\mS$ and all $B\in \mO_\mR$, by
\ba
\label{Ps}
P_\mS(A\otimes B)
:= \omega_\mR(B)\, A, 
\ea
and the same notation is used for its extension to $\tilde\mO$. Moreover,
for $a\in\{L,R\}$, the two-parameter family of mappings
$T_{\mS,\lam}^t, T_{\mS,\lam,a}^t:\mO_\mS\to\mO_\mS$ with
$\lam, t\in\R$ and $a\in\{L,R\}$ are defined, for all $A\in\mO_\mS$, by
\ba
\label{T-0}
T_{\mS,\lam}^t(A)
&:=P_{\mS}[\tilde\tau_0^{-t}\circ\tilde\tau_{\lam}^t(A\otimes 1)],\\
\label{Ta-0}
T_{\mS,\lam,a}^t(A)
&:=P_{\mS}[\tilde\tau_0^{-t}\circ\tilde\tau_{\lam,a}^t(A\otimes 1)].
\ea
The van Hove NESS $\omega_{\mS,+}\in\mE(\mO_\mS)$ 
with density $\rho_{\mS,+}\in\mL(\fh_\mS)$ 
and the Davies generator $K_{H,a}: \mO_\mS\to\mO_\mS$ 
of subreservoir $a\in\{L,R\}$ are defined,
for all $A\in\mO_\mS$, by
\ba
\label{macro-ness}
\omega_{\mS,+}(A)
&:=\lim_{t\to\infty}\,\lim_{\lam\to 0}\,
\omega_\mS(T_{\mS,\lam}^{t/\lam^2}(A)),\\
\label{dav-1}
K_{H,a}(A)
&:=\frac{\rd}{\rd t}\Big|_{t=0} \, \lim_{\lam\to 0} T_{\mS,\lam,a}^{t/\lam^2}(A),
\ea
if the limits exist. Finally, the van Hove energy current observable
$\Phi_{\mS,a}\in\mO_\mS$ and its expectation value $J_{\mS,a}$
in the van Hove NESS  are given by
\ba
\Phi_{\mS,a}\label{T-0}
&:= K_{H,a}(H_\mS),\\
J_{\mS,a}
&:=\omega_{\mS,+}(\Phi_{\mS,a}).
\ea
\ed
\br
For all $A\in\mO_\mS$ and $t>0$, defining 
\ba
K_H(A)
:=\frac{\rd}{\rd t}\Big|_{t=0}\lim_{\lam\to 0} 
T_{\mS,\lam}^{t/\lam^2}(A),
\ea
and  using 
$K_H=\sum_{a\in\{L,R\}} K_{H,a}$ (see, for example, Spohn and Lebowitz
\cite{SL}), and the invariance of the van Hove NESS under the  time evolution generated by $K_H$, 
we get the first law of thermodynamics of the van Hove regime, 
\ba
\sum_{a\in\{L,R\}} J_{\mS,a}
=0.
\ea
Hence, as for the microscopic regime,  we set $J_\mS:=J_{\mS,L}$.
\er

We begin our analysis by constructing the van Hove NESS. In the proof
of the following theorem (and the subsequent one), we will make use of
the reservoir time correlation function $\psi_a^\beta:\R\to\C$ 
with $a\in\{L,R\}$ and $\beta\in\R$ defined  by
\ba
\label{psia}
\psi_a^\beta(t)
:=(\delta_{\mR,a},\ia\planck_\beta(h_a)\e^{\i th_a}\ias \delta_{\mR,a}).
\ea
Moreover, we will use $\eps_i$ and $\pi_i$ with $i=1,\ldots ,n_\mS$ which are
the simple eigenvalues and the corresponding eigenprojections of the sample Hamiltonian $h_\mS$, respectively, given in Lemma \ref{lem:SamH} of
Appendix \ref{app:spec}, and, for $a\in\{L,R\}$ and $i=1,\ldots ,n_\mS$, we set
\ba
\label{Omega-ai}
\Omega_{a,i}
:=\|\pi_i\iss\delta_{\mS,a}\|^2,
\ea
and the scalar product and the norm in $\fh_\mS$ are denoted as the ones in $\fh$. The NESS can then be characterized as follows.
\bt[Van Hove NESS]
\label{thm:ma-ness}
There exists a unique quasifree van Hove NESS $\omega_{\mS,+}\in\mE(\mO_\mS)$ whose density has the form
\ba
\label{MaDens}
\rho_{\mS,+}
=\frac12 \sum_{a\in\{L,R\}}\planck_{\beta_a}(h_\mS).
\ea
\et
\bprf
Let us introduce the two-parameter family of states 
$\omega_{\mS,\lam}^t\in\mE(\mO_\mS)$ with 
$t,\lam\in\R$ which, for all $A\in\mO_\mS$, is defined by
\ba
\label{omega-lt}
\omega_{\mS,\lam}^t(A)
&:=\omega_\mS(T_{\mS,\lam}^t(A))\nonumber\\
&\hspace{1mm}=\tilde \omega\circ\tilde\tau_\lam^t(A\otimes 1).
\ea
Lemma \ref{lem:p-i}, \eqref{tau-lt}, and \eqref{rho} then imply that
their two-point function can be written as
\ba
\label{o-lt-a}
\omega_{\mS,\lam}^t(a^\ast(f)a(g))
=\sum_{a\in\{\mS,\mR\}} F_a(\lam,t),
\ea
where, for fixed $f,g\in\fh_\mS$, the function $F_a: 
\R^2\to\C$ with $a\in\{\mS,\mR\}$ is defined by
\ba
\label{Fa}
F_a(\lam,t)
:=(\e^{\i th_\lam}\is g,\ia \density_a\ias \e^{\i th_\lam}\is f).
\ea
In order to study the limit for $\lam\to 0$ of 
$F_a(\lam,t/\lam^2)$ with fixed $t>0$, we apply the weak coupling theory summarized in Appendix \ref{app:davies} in a form suitable for the present theorem (and for Theorem \ref{thm:ma-2law} below). Its ingredients are specified as follows: $\mH:=\fh$, $P_0:=\is\iss$, 
$P_1:=\ir\irs$, 
$U^t:=\e^{t Z}$ with $Z:=\i h_0$ (satisfying $[U^t,P_0]=0$ for all 
$t\in\R$), $A:=\sum_{a\in\{L,R\}} A_a$, $A_a:=\i v_a$, 
$V_\lam^t:=\e^{\i th_\lam}$,  
$W_\lam^t:=\is\iss \e^{\i t h_\lam}\is\iss$ and 
$R_\lam^t:=\ir\irs \e^{\i t h_\lam}\is\iss$.
In order to simplify the verification of the assumptions  
of the weak coupling theory, we 
define the operator-valued function $A_{a,b,c}^\beta:\R^3\to\mL(\fh)$ with $a,b,c\in\{L,R\}$ and $\beta\in\R$ by 
\ba
\label{Aabc}
A_{a,b,c}^\beta(r,s,t)
&:=2U^rP_0A_aP_1B_b^\beta U^s P_1 A_c P_0 U^t\nonumber\\
&= -\tfrac12 \delta_{ab}\delta_{ac} \,\psi_a^\beta(s)\, 
(\e^{-\i t h_0}\delta_{\mS,a},\cdot\,)\,\e^{\i r h_0}\delta_{\mS,a},
\ea
where we set $B_b^\beta:=i_b^{} \planck_\beta(h_b) i_b^\ast$ and the 
reservoir time correlation function $\psi_a^\beta$ is 
given in \eqref{psia}.
Let us begin with the sample contribution.
For $a=\mS$, we can write \eqref{Fa} as 
\ba
\label{FS-1}
F_\mS(\lam,t)
=\tfrac12\,(U^{-t}W_\lam^tP_0i_\mS^{}g, U^{-t}W_\lam^tP_0i_\mS^{}f).
\ea
In order to apply assertion {\it (1)} of Theorem \ref{thm:davies} 
of Appendix \ref{app:davies} on each factor of the scalar product in \eqref{FS-1}, we verify the following  
three assumptions of Theorem \ref{thm:davies}. Assumption {\it (a)} is  
$\dim(\ran(P_0))=n_\mS<\infty$. Assumption  {\it (b)} is $P_0AP_0=0$ and $P_1AP_1=0$ which follows from \eqref{v}.
It remains to verify assumption {\it (c)} which reads
$\int_0^\infty\!\!\rd t\,\,\|P_0AP_1 U^t P_1 A P_0\|<\infty$. Since
$P_1=2\sum_{b\in\{L,R\}} B^0_b$, we have
\ba
\label{(c)-1}
P_0AP_1 U^t P_1 A P_0
&=\sum_{a,b,c\,\in\{L,R\}} A^0_{a,b,c}(0,t,0)\nonumber\\
&=-\frac12 \sum_{a\in\{L,R\}} \psi_a^0(t)\, 
(\delta_{\mS,a},\cdot\,)\, \delta_{\mS,a},
\ea
from which it follows that $\|P_0AP_1 U^t P_1 A P_0\|
\le\frac12 \sum_{a\in\{L,R\}}|\psi^0_a(t)|\le 1$. 
In order to analyze the temporal decay  of \eqref{psia}, we
proceed to the diagonalization of $h_a$ by using Lemma
\ref{lem:subres} of Appendix \ref{app:spec}. Switching to the energy space $\widetilde\fh_+=L^2([-1,1];\rd e)$ of $h_a$, we get, 
for all $\beta\in\R$ and $a\in\{L,R\}$, that
\ba
\label{psia-1}
\psi^\beta_a(t)
=\frac{2}{\pi}\int_{-1}^1\rd e\,\, (1-e^2)^{1/2}\, 
\planck_\beta(e)\, \e^{\i t e},
\ea
which, by symmetry, is independent of $a$. From the asymptotic analysis
of Lemma \ref{lem:rtc} of Appendix \ref{app:davies}  (or by noting that,
for the case $\beta=0$, we can write $\psi^0_a(t)=J_1(t)/t$, where $J_1$ is the first order Bessel function), we have, for $t\to\infty$, that
\ba
\psi^\beta_a(t)
=\mO(t^{-3/2}).
\ea
Therefore, assumption {\it (c)} is also satisfied and we
can apply assertion {\it (1)} of Theorem \ref{thm:davies}. 
This assertion implies that, for any fixed $t>0$, we get
\ba
\lim_{\lam\to0} F_\mS(\lam,t/\lam^2)
=(g,\rho_{\mS\mS}^tf),
\ea
where 
the operator $\rho_{\mS\mS}^t\in\mL(\fh_\mS)$ is defined,
for all $t\in\R$, by
\ba
\label{rhoSSt}
\rho_{\mS\mS}^t
:=\frac12\, \iss \e^{t (K^\natural)^\ast}\!\e^{t K^\natural}\is.
\ea
Here, $K^\natural=\sum_{a\in\{L,R\}} K_a^\natural\in\mL(\fh)$, where, for $a\in\{L,R\}$, the operator $K_a^\natural\in\mL(\fh)$ is the spectral average from Theorem \ref{thm:davies} of the Davies generator 
$K_a\in\mL(\fh)$ given by
\ba
K_a
&:=\int_0^\infty \rd t\,\, U^{-t} P_0 A_a P_1 U^t P_1 A_a P_0\nonumber\\
&=\sum_{b\in\{L,R\}}\int_0^\infty \rd t\,\, A^0_{a,b,a}(-t,t,0)\nonumber\\
&=-\frac12 \int_0^\infty \rd t\,\, \psi_a^0(t) (\delta_{\mS, a},\cdot\,)
\, \e^{-\i t h_0}\delta_{\mS,a}.
\ea
For the computation of $K_a^\natural$, we make use of the fact that, for any
$A\in\mL(\fh)$, we can write 
$A^\natural=\sum_{i=1}^{n_\mS} \is \pi_i\iss A\is \pi_i\iss$, 
where $h_\mS=\sum_{i=1}^{n_\mS}\eps_i \pi_i$ 
with $\pi_i:=(\varphi_i,\cdot\,)\varphi_i$ is the
spectral representation of the sample Hamiltonian whose simple eigenvalues $\eps_i$ and the corresponding orthonormal eigenfunctions 
$\varphi_i$ are given in Lemma \ref{lem:SamH} of Appendix \ref{app:spec}. Using this representation, we find
\ba
\label{KNat}
K^\natural_a
=-\frac12\sum_{i=1}^{n_\mS} \Psi^0_a(\i\eps_i)\,\Omega_{a,i}\,\is\pi_i\iss,
\ea
where $\Psi^\beta_a$ denotes the Laplace transform of 
$\psi^\beta_a$ and $\Omega_{a,i}$ is given in \eqref{Omega-ai}.
Hence, for all $t\in\R$, we immediately get
\ba
\label{ExpK}
\e^{t K^\natural_a}
=\ir\irs
+\sum_{i=1}^{n_\mS} 
\e^{-\tfrac12 t \Psi^0_a(\i\eps_i)\Omega_{a,i}}\,\is \pi_i\iss.
\ea
Using that 
$[K^\natural_a,K^\natural_b]
=[K^\natural_a,(K^\natural_b)^\ast]
=0$ for all $a,b\in\{L,R\}$ and
plugging \eqref{ExpK} and its adjoint into \eqref{rhoSSt}, we find that, 
for any $t>0$, the sample contribution has the form 
\ba
\label{rhoSSt-1}
\rho_{\mS\mS}^t
&=\frac12 \,\iss \prod_{a\in\{L,R\}}\e^{t (K^\natural_a)^\ast} \prod_{b\in\{L,R\}}\e^{t K^\natural_b}\is\nonumber\\
&=\frac12 \sum_{i=1}^{n_\mS} \prod_{a\in\{L,R\}} 
\e^{-t\Re[\Psi^0_a(\i\eps_i)] \Omega_{a,i}}\, \pi_i.
\ea
We next turn to the  reservoir contribution.
For $a=\mR$, we can write \eqref{Fa} as 
\ba
F_\mR(\lam,t)
=\sum_{a\in\{L,R\}}(R_\lam^tP_0i_\mS g, B_a^{\beta_a} R_\lam^tP_0 i_\mS f),
\ea
and 
$B_a^{\beta_a}\ge 0$ and $[B_a^{\beta_a},U^t]=0$ for all $t\in\R$ and all $a\in\{L,R\}$. In order to determine the limit for $\lam\to 0$ of 
$F_\mR(\lam,t/\lam^2)$ with fixed $t>0$, we apply assertion {\it (2)} of Theorem \ref{thm:davies}. To this end, we have to verify that  
$S:=\sum_{a,b,c\in\{L,R\}} S_{a,b,c}$ converges in norm, 
where 
\ba
\label{Sabc}
S_{a,b,c}
&:=-2\int_0^\infty\rd t\,\,\Re[P_0A_aP_1B_b^{\beta_b} U^t
P_1A_cP_0 U^{-t}]\nonumber\\
&= -\int_0^\infty\rd t\,\, \Re[A^{\beta_b}_{a,b,c}(0,t,-t)]\nonumber\\
&=\frac12\,\delta_{ab}\delta_{ac}\int_0^\infty\rd t\,\, \Re[\psi^{\beta_a}_a(t) 
(\e^{\i t h_0}\delta_{\mS,a},\cdot\,)\,\delta_{\mS,a}].
\ea
Using again Lemma \ref{lem:rtc}, we get
$
\|\Re[P_0A_aP_1B_b^{\beta_b} U^t
P_1A_cP_0 U^{-t}]\|
\le \tfrac14|\psi^{\beta_a}_a(t)|
=\mO(t^{-3/2})$
for $t\to\infty$.
Hence, we apply assertion {\it (2)} of Theorem \ref{thm:davies} which implies, for any fixed $t>0$, that
\ba
\lim_{\lam\to 0}F_\mR(\lam,t/\lam^2)
=(g, \rho_{\mS\mR}^t f),
\ea
where $\rho_{\mS\mR}^t\in\mL(\fh_\mS)$ is defined, for all $t\in\R$, by
\ba
\label{rhoSR}
\rho_{\mS\mR}^t
:=\int_0^t\rd s\,\, \iss \e^{s (K^\natural)^\ast}
S^\natural \e^{s K^\natural}\is.
\ea
Using the spectral representation of the sample Hamiltonian 
$h_\mS$ as above, we get
\ba
\label{SB-1}
S_{a,b,c}^\natural 
=\frac12\delta_{ab}\delta_{ac}\sum_{i=1}^{n_\mS}\Re[\Psi^{\beta_a}_a(\i\eps_i)]\,\Omega_{a,i}\, \is \pi_i\iss.
\ea
Plugging \eqref{ExpK} and \eqref{SB-1} into \eqref{rhoSR}, we find that,
for any $t>0$, the reservoir contributes as
\ba
\label{rhoSR-1}
\rho_{\mS\mR}^t
&=\frac12 \sum_{a\in\{L,R\}}\int_0^t\rd s\,\, 
\iss \prod_{b\in\{L,R\}}\e^{s (K^\natural_b)^\ast}
S^\natural_{a,a,a} \prod_{c\in\{L,R\}}\e^{s K^\natural_c}\is\nonumber\\
&=\frac12\sum_{i=1}^{n_\mS} \frac{\sum_{a\in\{L,R\}}
\Re[\Psi^{\beta_a}_a(\i\eps_i)]\Omega_{a,i} }
{\sum_{b\in\{L,R\}} \Re[\Psi^0_b(\i\eps_i)]\Omega_{b,i}}
\left[1-\prod_{c\in\{L,R\}}\e^{-t\Re[\Psi^0_c(\i\eps_i)]\Omega_{c,i}}\right] \pi_i.
\ea
The denominator in \eqref{rhoSR-1} is strictly positive due to Lemma \ref{lem:rtc} and Lemma \ref{lem:SamH} which yield
that, for $i=1,\ldots ,n_\mS$, we have
\ba
\label{RePsi}
\Re[\Psi^\beta_a(\i\eps_i)]
&= 2(1-\eps_i^2)^{1/2}\varrho_\beta(\eps_i),\\
\label{Omega-ai-1}
\Omega_{a,i}
&=\frac{2}{n_\mS+1}\,(1-\eps_i^2),
\ea
where $\eps_i=\eps(k_i)$ and $k_i=i\pi/(n_\mS+1)$, and both
expressions are independent of $a\in\{L,R\}$. Using \eqref{rhoSSt-1} and \eqref{rhoSR-1}, we then find the density \eqref{MaDens} since
\ba
\omega_{\mS,+}(a^\ast(f)a(g))
&=\lim_{t\to\infty}\lim_{\lam\to 0}
\omega_{\mS,\lam}^t(a^\ast(f)a(g))\nonumber\\
&=\lim_{t\to\infty}(g, [\rho_{\mS\mS}^t+\rho_{\mS\mR}^t]f)\nonumber\\
&=(g,\rho_{\mS,+}f).
\ea
Moreover, it follows from the quasifreeness of the initial state and Lemma
\ref{lem:p-i} that the van Hove NESS is again quasifree. 
This is the assertion.
\eprf

Now we are able to determine the energy current expectation in the van Hove NESS or the van Hove entropy production given by
(see, for example, Aschbacher {\it et al.} \cite{AJPP1})
\ba
\Ep_\mS
:=-\sum_{a\in\{L,R\}} \beta_a J_{\mS,a}
=2\delta J_\mS.
\ea
In the following, $\tr$ denotes the trace over $\fh_\mS$.
\bt[Van Hove second law of thermodynamics]
\label{thm:ma-2law}
The van Hove entropy production  has the form
\ba
\label{JSa-1}
\Ep_\mS
=\frac{2\delta}{n_\mS+1}\, \tr[S_0(h_\mS)],
\ea
where $S_0$ is given in Theorem \ref{thm:2nd-order}.
Hence, if the system is truly out of equilibrium and $n>0$, the van Hove entropy production is strictly positive.
\et
\bprf
The van Hove energy current observable is given in Definition \ref{def:macro} by 
\ba
\Phi_{\mS,a}
&=K_{H,a}(H_\mS)\nonumber\\
&=\frac{\rd}{\rd t}\Big|_{t=0}
\lim_{\lam\to 0} T_{\mS,\lam,a}^{t/\lam^2}(H_\mS).
\ea
Moreover, for $n\in\N$, we know from \eqref{Hs1}  that the sample
 Hamiltonian has the form
\ba
\label{Hs1-1}
H_\mS
=\frac12 \sum_{x\in\Z_\mS'} 
[a^\ast(\iss\delta_x) a(\iss\delta_{x+1})+
a^\ast(\iss\delta_{x+1}) a(\iss\delta_x)],
\ea
where we set $\Z_\mS':=\Z_\mS\setminus\{n\}$.
For $n=0$, we have $H_\mS=0$ since $h_\mS=0$. 
Let us first consider \eqref{Ta-0} on the observable $A=a^\ast(f)a(g)$ 
for any $f,g\in\fh_\mS$. We then get
\ba
\label{T-1}
T_{\mS,\lam,a}^t(a^\ast(f)a(g))
=\sum_{b\in\{\mS,\mR\}} G_{b,a}(\lam,t),
\ea
where, for fixed $f,g\in\fh_\mS$, the map $G_{b,a}:\R^2\to\mO_\mS$ with 
$b\in\{\mS,\mR\}$ and  $a\in\{L,R\}$ is defined by
\ba
\label{As}
G_{\mS,a}(\lam,t)
&:=a^\ast(f_{\mS,a}(\lam,t))a(g_{\mS,a}(\lam,t)),\\
\label{Ar}
G_{\mR,a}(\lam,t)
&:=\omega_\mR(a^\ast(f_{\mR,a}(\lam,t))a(g_{\mR,a}(\lam,t)))\,
1_\mS,
\ea
and, for any $f\in\fh$, the function $f_{b,a}:\R^2\to\fh_b$ with $b\in\{\mS,\mR\}$ and $a\in\{L,R\}$ is given by
\ba
f_{b,a}(\lam,t)
:=i_b^\ast \e^{-\i t h_0}\e^{\i t h_{\lam,a}}\is f.
\ea
In order to study the limit $\lam\to 0$ of $G_{b,a}(\lam,t/\lam^2)$ 
for fixed $t>0$, we again apply the weak coupling theory from Appendix \ref{app:davies} with similar ingredients as in the proof of Theorem 
\ref{thm:ma-ness},
namely, $\mH:=\fh$, $P_0:=\is\iss$, $U^t:=\e^{tZ}$ with
$Z:=\i h_0$, $A_a:=\i v_a$, $V_{\lam,a}^t:=\e^{\i t h_{\lam,a}}$,
$W_{\lam,a}^t:=\is\iss \e^{\i t h_{\lam,a}} \is\iss$, and
$R_{\lam,a}^t:=\ir\irs \e^{\i t h_{\lam,a}} \is\iss$.
Let us start with the sample contribution \eqref{As}.
Since $\|a(f)\|=\|f\|$ for all $f\in\fh_\mS$,
it is enough  to study the weak coupling limit of 
\ba
f_{\mS,a}(\lam,t)
=\iss U^{-t} W_{\lam,a}^tP_0\is f.
\ea
The assumptions {\it (a)}, {\it (b)}, and {\it (c)} of Theorem 
\ref{thm:davies} are again verified as  in the proof of Theorem 
\ref{thm:ma-ness} with, in particular,
$\|P_0A_a P_1U^t P_1 A_aP_0\|\le \tfrac12 |\psi^0_a(t)|=\mO(t^{-3/2})$ for
$t\to\infty$, where here and in the following, we use the same notations as in the proof of Theorem \ref{thm:ma-ness}. It then follows from assertion {\it(1)} of Theorem \ref{thm:davies} and \eqref{ExpK} that,
for $t>0$, we have
\ba
\label{limS}
\lim_{\lam\to 0} f_{\mS,a}(\lam,t/\lam^2)
&=\iss \e^{t K_a^\natural}P_0\is f\nonumber\\
&=\sum_{i=1}^{n_\mS}\e^{-\frac12 t \Psi_a^0(\i\eps_i)\,\Omega_{a,i}}\,
\pi_i f.
\ea
We next turn to the reservoir contribution \eqref{Ar}.
In this case, we have to study the weak coupling limit of  
\ba
G_{\mR,a}(\lam,t)
=\sum_{b\in\{L,R\}}(R_{\lam,a}^t P_0 \is g, 
B^{\beta_b}_b R_{\lam,a}^t P_0 \is f) 1_\mS.
\ea
Since the additional assumption in assertion {\it(2)} of Theorem 
\ref{thm:davies} about the norm convergence of \eqref{SB}
is satisfied due to \eqref{Sabc} and the line following
it, we get from \eqref{ExpK} and \eqref{SB-1} that, for $t>0$, 
\ba
\label{limR}
\lim_{\lam\to 0}G_{\mR,a}(\lam,t/\lam^2)
&=\sum_{b\in\{L,R\}} \int_0^t \rd s\,\, (\e^{s K_a^\natural}P_0\is g,
S_{a,b,a}^\natural \e^{s K_a^\natural}P_0\is f)\,1_\mS\nonumber\\
&=\frac12\sum_{i=1}^{n_\mS}\frac{\Re[\Psi^{\beta_a}_a(\i\eps_i)]}
{\Re[\Psi^0_a(\i\eps_i)]}
\left[1-\e^{-t\Re[\Psi^0_a(\i\eps_i)]\Omega_{a,i}}\right](g,\pi_i f)\,
1_\mS.
\ea
Therefore, from \eqref{limS} and \eqref{limR}, we find
\ba
\label{K-1}
K_{H,a}(a^\ast(f)a(g))
&=\sum_{b\in\{\mS,\mR\}} \frac{\rd}{\rd t}\Big|_{t=0}\,
\lim_{\lam\to 0} G_{b,a}(\lam,t/\lam^2)\nonumber\\
&=-\frac12 \sum_{i,j=1}^{n_\mS} \big[\Psi^0_a(\i\eps_i)\Omega_{a,i}
+ \bar\Psi^0_a(\i\eps_j)\Omega_{a,j}\big]\,
a^\ast(\pi_i f)\, a(\pi_j g)\nonumber\\
&+ \frac12\sum_{i=1}^{n_\mS}\Re\big[\Psi^{\beta_a}_a(\i\eps_i)\big]\Omega_{a,i}
(g,\pi_i f)\,1_\mS.
\ea
Applying \eqref{K-1} to \eqref{Hs1-1}, plugging the resulting
expression into the van Hove NESS from Theorem \ref{thm:ma-ness}, and using \eqref{RePsi}, \eqref{Omega-ai-1}, and 
$\sum_{x\in\Z_\mS'} \Re[(\iss\delta_x, \pi_i\iss\delta_{x+1})]=\eps_i$,
we arrive at the assertion.
\eprf

We finally get the following result.
\bt[Van Hove is second order]
\label{thm:mima}
The van Hove entropy production is the leading second order contribution to the small coupling expansion of the microscopic 
entropy production, 
\ba
\Ep_\mS
=\Ep.
\ea
\et

\vspace{3mm}

\bprf
Due to Lemma \ref{lem:SamH} in Appendix \ref{app:spec} which states that
the eigenvalues of the sample Hamiltonian have the form 
$\eps_i=\eps(k_i)$ for all $i=1,\ldots ,n_\mS$, we immediately get the assertion by comparing \eqref{Ja2} in Theorem \ref{thm:2nd-order} and \eqref{JSa-1} in Theorem \ref{thm:ma-2law}.
\eprf
\br
In Aschbacher {\it et al.} \cite{AJPP1}, an assertion like
Theorem \ref{thm:mima} has been derived  for the simple electronic black box model (SEBB) with one-dimensional sample system. The assumptions made there on the 
SEBB model compare to Lemma \ref{lem:coupled} and \ref{lem:SamH} of Appendix \ref{app:spec}, Definition \ref{def:qfs}, and Lemma \ref{lem:rtc} of Appendix \ref{app:davies}.
\er
\br
In Aschbacher and Spohn \cite{AS}, a simple sufficient condition has been
established which ensures the strict positivity of the entropy production
as soon as the microscopic regime is related to the van Hove regime
as in Theorem \ref{thm:mima}. In order to be able to apply this criterion,
one assumption on the so-called effective coupling and another one on the 
triviality of some commutants have to be satisfied. Whereas it has been shown in \cite{AS} that the entropy production can still be strictly
positive if the latter condition is violated, the present case is an
example showing that the criterion is not necessary due to violation of the former condition. In order to formulate this condition
precisely, we rewrite the couplings \eqref{V1a} as
$\tilde V_a
=\sum_{i=1}^2 V_{\mS,a,i}^{(1)}\otimes 
V_{\mR,a,i}^{(1)}$, where $V_{\mS,a,1}^{(1)}
=\Re[a(\iss\delta_{\mS,a}) \Gamma(-1)]$, 
$V_{\mS,a,2}^{(1)}
=-\Im[a(\iss\delta_{\mS,a}) \Gamma(-1)]$, and 
\ba
V_{\mR,a,1}^{(1)}
&=\Re[a(\irs\delta_{\mR,a})],\\
V_{\mR,a,2}^{(1)}
&=\Im[a(\irs\delta_{\mR,a})].
\ea
Moreover, we define the matrix-valued reservoir correlation function
$R_a: \R\to\C^{2\times 2}$ by
\ba
\label{Raij}
R_{a,ij}(t)
:=\omega_\mR(\tau_\mR^t(V_{\mR,a,i}^{(1)})V_{\mR,a,j}^{(1)}).
\ea
The effective coupling conditions from \cite{AS} then requires that,
for all $a\in\{L,R\}$ and for all energies 
$\eps\in\spec(H_\mS)-\spec(H_\mS)$,
the temporal Fourier transform of the reservoir correlation matrix should be positive definite, 
\ba
\label{ecc}
\hat R_a(\eps)
>0.
\ea
Now, due to \eqref{Hs1}, and \eqref{hS-ev} from Appendix \ref{app:spec}, we have on 
$\fF(\fh_\mS)=\C\oplus(\oplus_{\alpha=1}^{n_\mS}\fh_\mS^{\wedge\alpha})$
that
\ba
\spec(H_\mS)
=\{0\}\cup \left(\bigcup_{\alpha=1}^{n_\mS} 
\{\eps_{i_1}+\ldots +\eps_{i_\alpha}\}_{1\le i_1<\ldots <i_\alpha\le n_\mS}\right),
\ea
where $\eps_i$ with $i=1,\ldots ,n_\mS$ are the eigenvalues of the one-particle sample Hamiltonian $h_\mS$. On the other hand,
due to the specific form of the Planck density, 
we have 
from \eqref{rtc-2} of Appendix \ref{app:davies} that
$
\hat\psi_a^\beta(-e)
=\hat\psi_a^{-\beta}(e)
=\e^{\beta e} \hat\psi_a^\beta(e)$ 
which, together with \eqref{Raij}, yields 
\ba
\hat R_{a,ij}
=\frac{\delta_{ij}}{2}\, \hat\psi_a^{\beta_a}.
\ea
Choosing  $n=1$, we have 
$\eps_1=-\eps_3=\sqrt{2}/2$, and, hence,
$\eps_1-\eps_3>1$. Since $\hat\psi_a^{\beta_a}(\eps_1-\eps_3)=0$ due
to Lemma \ref{lem:rtc} in Appendix \ref{app:davies}, the effective coupling condition \eqref{ecc} 
is not satisfied for all energy differences.
\er
\br
As we indicated repeatedly in the appendix, the derivation of Theorem \ref{thm:mima}
for the full anisotropic XY model with an additional external magnetic field is much more complicated. This is also true for the derivation of 
a theorem like Theorem \ref{thm:mima} for the isotropic case and 
 general observables.
We will study these question for more general quasifree systems 
elsewhere.
\er
\begin{appendix}
\section{Spectral properties}
\label{app:spec}

In this section, we display some spectral properties of the different
Hamiltonians appearing in the model. In the first lemma, we introduce what we will call the energy space 
of the XY Hamiltonian  $h$ being the
direct integral decomposition of the absolutely continuous subspace 
 w.r.t. which $h$ is diagonal, namely
\ba
\wt\fh
:=L^2([-1,1],\C^2;\rd e).
\ea
Moreover, the map $\wt \ff\in\mL(\wh\fh,\wt\fh)$ is defined, for all
$\varphi\in\wh\fh$, by
\ba
\label{ftilde}
\wt\ff \varphi (e)
:=
(2\pi)^{-1/2}(1-e^2)^{-1/4}\,
[\varphi(\arccos(e)), \varphi(-\arccos(e))],
\ea
and the momentum space 
$\widehat\fh=L^2([-\pi,\pi];\tfrac{\rd k}{2\pi})$
has been introduced in
\eqref{h-hat}. We will use the notation 
$\wt f:=\wt\ff\ff f$ for all $f\in\fh$, and 
$\wt A:= \wt\ff\ff A \ff^\ast\wt\ff^\ast$ 
for all $A\in\mL(\fh)$, where the Fourier transform 
$\ff:\fh\to\wh\fh$ is also given after  \eqref{h-hat}.
Moreover, the Euclidean scalar product in the fiber $\C^2$ is denoted by
$\langle\,\cdot\,,\,\cdot\rangle_2$.
\bl[XY Hamiltonian]
\label{lem:xy}
The XY Hamiltonian $h\in\mL(\fh)$ has purely absolutely continuous
spectrum with $\spec(h)=[-1,1]$, and it is diagonal in $\wt\fh$.
\el
\bprf
In momentum space $\wh\fh$, the Hamiltonian $\wh h$ acts as the 
multiplication by the dispersion relation $\eps(k)$ from Theorem 
\ref{thm:ness}. Moreover, a simple computation shows that 
${\wt\ff}$ is a surjective 
isometry with
${\wt\ff}^{-1}=\wt\ff^\ast:\wt\fh\to\wh\fh$ 
acting on all $\eta=:[\eta_1,\eta_2]\in\wt\fh$ as
\ba
\label{ftilde-ast}
\wt\ff^\ast\eta (k)
=(2\pi)^{1/2}(1-\eps^2(k))^{1/4}\,
[\chi_{[0,\pi]}(k)\,\eta_1(\eps (k))
+\chi_{[-\pi,0]}(k)\,\eta_2(\eps (k))].
\ea
This implies the assertion.
\eprf
\br
For $\gam\neq0$, the energy space for the
XY Hamiltonian, now acting on $\fh^{\oplus 2}$ (see Remark \ref{rem:XY}),
takes the form
$L^2([-1,-|\gam|],\C^4;\rd e)\oplus L^2([|\gam|,1],\C^4;\rd e)$ (and additional $\C^2$-valued factors if $\mu\neq 0$). Moreover, the 
nondiagonal matrix-multiplication operator by which its Fourier transform
acts in momentum space $L^2([-\pi,\pi];\tfrac{\rd k}{2\pi})^{\oplus 2}$
has to be diagonalized.
\er

The subreservoir Hamiltonians have similar properties. Let us introduce the spaces $\fh_+:=\ell^2(\N)$ and
$\widehat\fh_+:=L^2([0,\pi]; \tfrac2\pi\rd k)$, and 
\ba
\widetilde\fh_+
:=L^2([-1,1];\rd e).
\ea
\bl[Subreservoir Hamiltonians]
\label{lem:subres}
The subreservoir Hamiltonians 
$h_a\in\mL(\fh_a)$ with $a\in\{L,R\}$
have purely absolutely continuous
spectrum with $\spec(h_a)=[-1,1]$, and they are diagonal in 
$\widetilde\fh_+$.
\el
\bprf
We use the unitary mappings $\fh_a
\stackrel{t_a}{\to}\fh_+
\stackrel{\fs}{\to}\widehat\fh_+
\stackrel{\wt \fs}{\to}\widetilde\fh_+$, where the ingredients are
given by $t_Lf(x):=f(-(x+n))$ and $t_Rf(x):=f(x+n)$ for all 
$f\in \fh_a$, by the Fourier-sine transform 
$\fs(f)(k):=\sum_{x=1}^\infty f(x)\sin(xk)$ for all $f\in \fh_+$, and by the energy transformation which, for all 
$\varphi\in \widehat\fh_+$, has the form
\ba
\wt \fs\varphi(e)
:=2^{1/2}\pi^{-1/2} (1-e^2)^{-1/4}\, \varphi(\arccos(e)).
\ea
In $\widehat\fh_+$, the subreservoir Hamiltonians act
by multiplication with $\eps(k)$, and applying the energy transformation, we get the assertion.
\eprf

Next, we turn to the  coupled and decoupled Hamiltonians. 
We denote by $\spec_{\rm sc}(A)$, 
$\spec_{\rm ac}(A)$, and $\spec_{\rm pp}(A)$ the singular continuous, 
the absolutely continuous, and the pure point spectrum of the operator 
$A$, respectively.
\bl[{[}De{]}coupled Hamiltonian]
\label{lem:coupled}
For all $\lam\in\R$, it holds that
 $\spec_{\rm sc}(h_\lam)=\emptyset$ and 
$\spec_{\rm ac}(h_\lam)=[-1,1]$. Moreover, the coupled Hamiltonian 
has the properties $\spec_\rpp(h_\lam)=\emptyset$ for all 
$0<|\lam|\le 1$
and ${\rm card}(\spec_\rpp(h_\lam))\le 2$ for all $|\lam|>1$. The 
decoupled Hamiltonian satisfies ${\rm card}(\spec_\rpp(h_0))= n_\mS$.
\el
\bprf
For the first two assertions and the fact that 
${\rm card}(\spec_{\rm pp}(h_\lam))<\infty$ for all $\lam\in\R$, see, for example, Hume and Robinson \cite{HR} (in fact, this is all what is used in Theorem \ref{thm:ness-2}). Next, let $\lam\in\R\setminus\{0\}$, and 
let us assume that there exist $0\neq f\in\fh$ and $e\in[-1,1]$ s.t.
$h_\lam f=e f$. Written out and evaluated at any $x\in\Z$, this equation
reads
\ba
\label{eve}
f(x+1)+f(x-1)+(\lam-1)\sum_{i=1}^4(\delta_i^1,f)\,\delta_i^2(x)
=2e f(x),
\ea
where $\delta^1, \delta^2\in\fh^4$ are given in Lemma \ref{lem:Rb}
of Appendix \ref{app:wave}.
It follows from \cite{HR} that eigenfunctions corresponding to eigenvalues in $[-1,1]$ satisfy $f(x)=0$ for all $|x|\ge n+1$. Hence,
plugging  $x=\pm(n+1)$ into \eqref{eve}, we find that $f(\pm n)=0$. The
eigenvalue equation then becomes $h_\lam f=hf=e f$ which leads to 
$\spec_{\rpp}(h_\lam)\cap[-1,1]=\emptyset$ for all  
$|\lam|>0$. Let us next consider the eigenvalue
equation $h_\lam f=e f$ for $0\neq f\in\fh$ and $e\in\R$ with $|e|>1$
(see also Lemma \ref{lem:Rb} of Appendix \ref{app:wave}).
Plugging $x=\pm n, \pm (n+1)$ into \eqref{eve} and setting 
$f^n:=[f(-n), f(-n-1),f(n+1),f(n)]\in\C^4$, we get
\ba
\Sigma_\lam(e) f^n
=0,
\ea
where the matrix $\Sigma_\lam(e)\in\C^{4\times 4}$ is given in 
Lemma \ref{lem:Rb} and Lemma \ref{lem:res} of Appendix \ref{app:wave}. If 
$\det(\Sigma_\lam(e))\neq 0$, we again get $h_\lam f=hf=e f$. Hence, the
eigenvalues are the solutions of  $\det(\Sigma_\lam(e))=0$, where,
analogously to Proposition \ref{prop:inv} of Appendix \ref{app:wave}, 
we have
\ba
\label{eve-1}
\det(\Sigma_\lam)
=\prod_{\sigma=\pm 1}[(1-\lam^2) E^{2n+2}+\sigma (\lam^2 E^2-1)],
\ea
and $E(e)=e-\sign(e)(e^2-1)^{1/2}$ stems from \eqref{Esg} of Appendix
\ref{app:wave}. 
Using that $0< E^2(e)<1$ for all $e\in\R$ with $|e|>1$, none of the 
two factors in \eqref{eve-1} vanishes if $0<|\lam|\le 1$. On the other hand, if $|\lam|>1$, the factor with $\sigma=1$ has at most one root (depending
on the size of $|\lam|$, it may have no root for small $n$ but for
sufficiently large $n$, it has one root), whereas the 
factor with $\sigma=-1$ has exactly one root for all $n\in\N_0$.
Finally, for $\lam=0$, we know from Remark \ref{rem:dec}, that $h_0$ does
not couple the subsystems to each other. Using Lemma \ref{lem:subres} and Lemma \ref{lem:SamH}, we then arrive at the assertion.
\eprf

The spectral resolution of the sample Hamiltonian can be explicitly
determined.
\bl[Sample Hamiltonian]
\label{lem:SamH}
The spectrum of the sample Hamiltonian $h_\mS\in\mL(\fh_\mS)$ consists of
$n_\mS$ nondegenerate eigenvalues which, for $i=1,\ldots ,n_\mS$, have the form
\ba
\label{hS-ev}
\eps_i
=\eps(k_i),\quad
k_i
:=\frac{i\pi}{n_\mS+1}.
\ea
The corresponding orthonormal eigenfunctions $\varphi_i\in\fh_\mS$
are given, for all $x\in\Z_\mS$, by
\ba
\label{hS-ef}
\varphi_i(x)
=\left(\frac{2}{n_\mS+1}\right)^{1/2} 
\sin\left(\left[x+\frac{n_\mS+1}{2}\right]k_i\right).
\ea
\el
\bprf
Note that the sample Hamiltonian 
$h_\mS=\iss h\is\in\mL(\fh_\mS\simeq\C^{n_\mS})$ is the usual discrete
Laplacian acting by application of  the matrix $[h_\mS]_{ij}=\tfrac12 (\delta_{ij+1}+\delta_{ij-1})$ for $i,j=1,\ldots ,n_\mS$ (see, for example, B\"ottcher and Grudsky \cite{BG}).
\eprf
\section{Wave operator}
\label{app:wave}

In this section, we use the stationary approach to scattering theory in order to compute the wave operators appearing in the NESS expectation
value of the energy current observable. To this end,  we first express the resolvent of the coupled  Hamiltonian by the resolvent of the XY Hamiltonian. 
For any operator $A\in\mL(\mH)$, we denote by $\res(A)$ the resolvent set
of $A$ and by $r_z(A):=(A-z)^{-1}\in\mL(\mH)$ the resolvent of $A$ at the point $z\in\res(A)$.
\bl[Coupled resolvent]
\label{lem:Rb}
For $z\in\res(h)\cap\res(h_\lam)$, we have
\ba
\label{Rb}
r_z(h_\lam)
=r_z(h)-\frac{\lam-1}{2}\sum_{i,j=1}^4 \Sigma^{-1}_{\lam, ij}(z)\,
 (r_{\bar z}(h) \delta^1_j,\,\cdot\,)\, r_{z}(h) \delta^2_i,
\ea
where  $v=\frac12\sum_{i=1}^4
(\delta^1_i,\cdot\,)\,\delta^2_i$ and
$\delta^1:=[\delta^1_i]_{i=1}^4, \delta^2:=[\delta^2_i]_{i=1}^4\in \fh^4$ 
are given by
\ba
\label{fv}
\delta^1
&:=[\delta_{\mS,L}, \delta_{\mR,L}, \delta_{\mR,R},\delta_{\mS,R}],\\
\label{gv}
\delta^2
&:=[\delta_{\mR,L}, \delta_{\mS,L}, \delta_{\mS,R},\delta_{\mR,R}].
\ea
Moreover, the interaction matrix 
$\Sigma_\lam(z)\in\C^{4\times 4}$ is defined, for $i,j=1,\ldots ,4$, by
\ba
\label{Sigma-z}
\Sigma_{\lam, ij}(z)
:=\delta_{ij}+\frac{\lam-1}{2}\,\, (\delta^1_i, r_{z}(h) \delta^2_j).
\ea
\el
\br
For $\gam\neq0$, the wave operator in the selfdual setting acts on
$\fh^{\oplus 2}$ and has nonvanishing off-diagonal components (whereas
for $\gamma=0$ it is block-diagonal). The interaction matrix then lies
in $\C^{8\times 8}$.
\er
\bprf
In order to simplify the notation,  we drop
the indices of the resolvents. Using the resolvent identity $r(h_\lam)=r(h)-(\lam-1) r(h) v r(h_\lam)$,
we have, for all $f\in\fh$, 
\ba
\label{res-1}
r(h_\lam)f
+\frac{\lam-1}{2}\sum_{j=1}^4(\delta^1_j, r(h_\lam)f)\, r(h) \delta^2_j
=r(h)f.
\ea
Taking the scalar product of \eqref{res-1} with $\delta^1_i$ 
for all $i=1,\ldots ,4$, we get
\ba
\label{Aeq}
\left(1+\tfrac{\lam-1}{2} A\right)\xi=\eta,
\ea
 where the components of  
$\xi,\eta\in\C^4$ and $A\in\C^{4\times 4}$ are defined, for 
$i,j=1,\ldots ,4$, by
\ba
\xi_i
&:= (\delta^1_i, r(h_\lam)f),\\
\eta_i
&:= (\delta^1_i, r(h)f),\\
A_{ij}
&:= (\delta^1_i, r(h) \delta^2_j).
\ea
Moreover, defining $B\in\C^{4\times 4}$ by 
$B_{ij}:=(\delta^1_i,r(h_\lam) \delta^2_j)$ for $i,j=1,\ldots ,4$, the resolvent identity implies that, for any $\lam\in\R$, we have
$(1+\tfrac{\lam-1}{2} A)(1-\tfrac{\lam-1}{2} B)=1$, so 
$1+\tfrac{\lam-1}{2} A$ is invertible. We now solve \eqref{Aeq} for $\xi$ and plug the resulting expression into \eqref{res-1}. This yields
the assertion.
\eprf

We next introduce the following abbreviations.
\bd[Boundary values]
\label{def:boundary}
Let $z\in\res(h)$, $e\in\R$, $\veps>0$, and $f,g\in\fh$. We define
\ba
\label{rho-fg}
\varrho_{f,g}(z)
&:=(f,r_{z}(h)g),\\
\label{gamma-fg}
\gamma_{f,g}(e,\veps)
&:=\frac{1}{2\pi\i}
\left(\varrho_{f,g}(e+\i\veps)-\varrho_{f,g}(e-\i\veps)\right),
\ea
and, if the limits exist, we write
\ba
\label{rho-0}
\varrho_{f,g}(e\pm\i 0)
&:=\lim_{\veps\to 0^+} \varrho_{f,g}(e\pm\i\veps),\\
\label{gamma-0}
\gamma_{f,g}(e)
&:=\lim_{\veps\to 0^+} \gamma_{f,g}(e,\veps).
\ea
\ed

Let us recall from Lemma \ref{lem:xy} of Appendix \ref{app:spec} that 
$\wt\fh=L^2([-1,1],\C^2;\rd e)$ is the energy space of the XY Hamiltonian
$h$. The wave operator then looks as follows.
\bp[Wave operator]
\label{prop:wave-diag}
In the energy space $\wt\fh$ of the XY Hamiltonian, the action of the wave operator is given, for all $f\in\fh$, by
\ba
\label{wave-diag}
\wt w(h,h_\lam)\wt f
=\wt f
-\frac{\lam-1}{2}\sum_{i,j=1}^4 \Sigma_{\lam, ij}^{-1}(\,\cdot-\i 0)\,
\varrho_{\delta^1_j,f}(\,\cdot-\i0)
\,\wt \delta^2_i,
\ea
where, for all $e\in(-1,1)$, the boundary interaction matrix
$\Sigma_\lam(e-\i 0)\in\C^{4\times 4}$ is defined by 
\ba
\label{mat-0}
\Sigma_{\lam, ij}(e-\i 0)
:=\delta_{ij}+\frac{\lam-1}{2}\,
\varrho_{\delta^1_i,\delta^2_j}(e-\i 0).
\ea
\ep
\bprf
In order to compute the wave operator with the help of stationary scattering theory, we rewrite it in its weak abelian form (see,
for example, Yafaev \cite{Y}),
\ba
\label{wave0}
w(h,h_\lam)
=\wlim_{\veps\to 0^+} \, 2\veps \int_0^\infty\rd t\,\,
{\rm e}^{-2\veps t} 1_{\rm ac}(h) {\rm e}^{-{\rm i}th}
{\rm e}^{{\rm i}th_\lam}1_{\rm ac}(h_\lam).
\ea
Applying Parseval's identity to \eqref{wave0},
and using that 
$r_{e-\i\veps}(h)=-\i\int_0^\infty\rd t\,\,
\e^{\i t(h-(e-\i\veps))}$, we get, for all $f,g\in\fh$, that
\ba
\label{wave1}
(f,w(h,h_\lam)g)
=\lim_{\veps\to 0^+}\frac{\veps}{\pi} \int_{-\infty}^\infty\rd e\,\,
(r_{e-\i\veps}(h)1_{\rm ac}(h)f, r_{e-\i\veps}(h_\lam)
1_{\rm ac}(h_\lam)g).
\ea
Moreover, if the limit $\veps\to 0^+$ of 
$\veps (r_{e-\i\veps}(h)f, r_{e-\i\veps}(h_\lam)g)$
exists for all $f,g\in\fh$ and almost all $e\in\R$ (the set of full measure depending on $f$ and $g$) and using that 
$1_{\rm ac}(h)=1$ and $\spec(h)=[-1,1]$, we can write
\ba
\label{wave2}
(f,w(h,h_\lam)g)
=\int_{-1}^1\rd e\,\,
\lim_{\veps\to 0^+}\frac{\veps}{\pi} 
(r_{\e-\i\veps}(h)f, r_{e-\i\veps}(h_\lam)g).
\ea
In order to compute the  limit in \eqref{wave2}, we express the  resolvent $r_{e-\i\veps}(h_\lam)$ of the coupled Hamiltonian in terms of 
the resolvent $r_{e-\i\veps}(h)$ of the XY Hamiltonian.  Plugging 
\eqref{Rb} into the scalar product on the r.h.s. of \eqref{wave2}, we have
\ba
\label{integrand1}
\frac{\veps}{\pi}\,
(r_{e-\i\veps}(h)f,r_{e-\i\veps}(h_\lam)g)
=\gamma_{f,g}(e,\veps)
-\frac{\lam-1}{2}\sum_{i,j=1}^4 \gam_{f,\delta^2_i}(e,\veps)\,
 \Sigma_{\lam, ij}^{-1}(e-\i\veps)\,\varrho_{\delta^1_j,g}(e-\i\veps),
\ea
where  we used that 
$\frac{\veps}{\pi}(r_{e-\i\veps}(h)f,r_{e-\i\veps}(h)g)=\gamma_{f,g}(e,\veps)$ (which follows from the resolvent identity). Now, we know that, for all $f,g\in\fh$ and almost all $e\in[-1,1]$, the limit
\ba
\varrho_{f,g}(e\pm\i 0)
\label{rho-1}
=\pm\pi \i \frac{\rd  (f,\zeta(e)g)}{\rd e}\,
+{\rm p.v.}\!\int_{-1}^1 \rd e'\, \frac{1}{e'-e} \,
\frac{\rd (f, \zeta(e')g)}{\rd e'}
\ea
exists, where the ${\rm p.v.}$-integral is Cauchy's principle value,  the mapping $\zeta:\mB(\R)\to\mL(\fh)$ denotes the projection-valued spectral measure of the XY Hamiltonian $h$ with $\mB(\R)$ the Borel sets on $\R$,
and we used that
\ba
\label{supp}
\rd(f, \zeta(e)g)
=\chi_{[-1,1]}(e)\, \frac{\rd (f, \zeta(e)g)}{\rd e}\,\,  \rd e.
\ea
Moreover, it follows from \eqref{gamma-fg} and \eqref{rho-1} that
\ba
\label{gamma-0}
\gamma_{f,g}(e)
=\frac{\rd  (f,\zeta(e)g)}{\rd e}.
\ea
Hence, we find that
\ba
\label{wave3}
(f,w(h,h_\lam)g)
=(f,g)
-\frac{\lam-1}{2} \sum_{i,j=1}^4 
\int_{-1}^1\rd e\,\, \gam_{f,\delta^2_i}(e)\,
\Sigma_{\lam, ij}^{-1}(e-\i 0)\,
\varrho_{\delta^1_j,g}(e-\i 0),
\ea
where the invertibility of the interaction matrix is assured as in 
the proof of Lemma \ref{lem:Rb} and we used 
$\int_{-1}^1\rd e\,\, \gamma_{f,g}(e)=(f,1_{\rm ac}(h)g)=(f,g)$
in the first term on the r.h.s. of \eqref{wave3}. In order to write the derivatives in \eqref{gamma-0} entering \eqref{wave3} more explicitly, we switch to the energy space representation using $\wt\fh$ of Lemma \ref{lem:xy}. This lemma implies that
\ba
\label{derivative}
\frac{\rd  (f,\rho(e)g)}{\rd e}
={\langle \wt f(e),\wt g(e)\rangle}_2,
\ea
where we recall that ${\langle\cdot,\cdot\rangle}_2$ 
denotes the Euclidean scalar product in the fiber $\C^2$ of the direct integral $\wt\fh$, and $\wt f=\wt\ff\ff f$ for all 
$f\in\fh$. Hence, plugging  \eqref{gamma-0} and 
\eqref{derivative} into \eqref{wave3}, we arrive at the assertion.
\eprf

In order to completely determine the wave operator, we have to compute the
boundary values and the inverse of the interaction matrix. To this
end, we define the function\linebreak $E:\R\to\C$ by 
\ba
\label{Esg}
E(e)
:=\begin{cases}
e+\i(1-e^2)^{1/2}, & \mbox{if $|e|\le 1$},\\
e-\sign(e)(e^2-1)^{1/2}, & \mbox{if $|e|>1$}.
\end{cases}
\ea
Let us start with the computation of some XY
resolvent amplitudes for completely localized wave functions (which is
also used in Appendix \ref{app:spec}).
\bl[Resolvent amplitudes]
\label{lem:res}
For $x\in\Z$, we have
\ba
\label{rho00}
-\frac{2E^{|x|+1}(e)}{1-E^2(e)}
=\begin{cases}
\varrho_{\delta_0,\delta_x}(e-\i 0), & \mbox{if \,$|e|<1$},\\
\varrho_{\delta_0,\delta_x}(e),      & \mbox{if \,$|e|>1$}.
\end{cases}
\ea
\el
\bprf
Let $x\in\Z$ with $x\ge 0$, $e\in (-1,1)$, and $\varepsilon>0$ sufficiently small. We first rewrite the momentum space representation
of the resolvent amplitude in the form of a contour integral over the positively oriented unit circle $\T$ as 
\ba
\label{rho00-1}
\varrho_{\delta_0,\delta_x}(e-\i \veps)
=\frac{1}{\i\pi}\oint_{\T}\rd z\,\,\frac{z^x}
{z^2-2(e-\i\veps)z+1}.
\ea
Then, using Cauchy's residue theorem and taking the limit $\eps\to 0^+$, we get the first expression in
\eqref{rho00} for $x\ge 0$.
Moreover, due to
the parity invariance of the XY Hamiltonian, $[h,\theta]=0$, 
where $\theta:\fh\to\fh$ is defined, for all
$f\in\fh$, by $(\theta f)(x):=f(-x)$, 
we also have 
$\varrho_{\delta_0,\delta_{-x}}(e-\i \eps)
=\varrho_{\delta_0,\delta_{x}}(e-\i \eps)$.
The second assertion is derived similarly. 
\eprf
\br
For $\gam\neq 0$ and $\mu=0$, one gets an analogous expression for 
\eqref{rho00-1} but, in this case, there is a nontrivial 
numerator, and the polynomial in the denominator becomes biquadratic,
$z^4+az^2+1$, where $a$ depends on $\gam$, $e$, and $\veps$. 
Moreover, if both $\gam\neq 0$ and $\mu\neq 0$, this polynomial changes to
$z^4+az^3+bz^2+az+1$, where $a$ depends on $\mu$ and $\gam$,
and $b$ on $\mu$, $\gam$, $e$, and $\veps$. Hence, the computation of
the roots becomes increasingly and substantially more complicated (see also, for example,  Carey and Hume \cite{CH}).
\er

We next turn to the computation of the inverse of the boundary value
interaction matrix from Proposition \ref{prop:wave-diag}. 
For the convenience of the reader who wants to work with this 
nonequilibrium model and for reasons of a possible future extension, we 
display the detailed results of the computations.
\bp[Inverse boundary interaction matrix]
\label{prop:inv}
For all $e\in(-1,1)$, we have
\ba
\label{inv-0}
\Sigma_{\lam}^{-1}(e-\i 0)
=\frac{1}{\Delta_{\lam}(e)}
\begin{bmatrix}
M_{\lam}(e) & N_{\lam}(e) \\
\sigma_1 N_{\lam}(e) \sigma_1 & \sigma_1 M_{\lam}(e) \sigma_1 
\end{bmatrix}.
\ea
Here, the determinant $\Delta_{\lam}:=\det[\Sigma_{\lam}(\,\cdot-\i0)]: (-1,1)\to \C$ reads
\ba
\label{Delta-1}
\Delta_{\lam}
=\frac{1-E^{4n+4}}{(1-E^2)^2}
-\frac{2E^2(1-E^{4n+2})}{(1-E^2)^2}\,\lam^2
+\frac{E^4(1-E^{4n})}{(1-E^2)^2}\,\lam^4,
\ea
the matrix-valued functions $M_{\lam}, N_{\lam}:(-1,1)\to\C^{2\times 2}$ have the structure
\ba
\label{inv-1}
M_{\lam}
:=\begin{bmatrix}
a_{\lam} & b_{\lam}\\
c_{\lam} & a_{\lam}
\end{bmatrix},\quad
N_{\lam}
:=\begin{bmatrix}
d_{\lam} & e_{\lam}\\
e_{\lam} & f_{\lam}
\end{bmatrix},
\ea
and the functions $a_{\lam},\ldots , f_{\lam}: (-1,1)\to \C$ are defined by
\ba
a_{\lam}
&:=\frac{1-E^{4n+4}}{(1-E^2)^2}
-\frac{E^2(1-E^{4n+2})}{(1-E^2)^2}\,(\lam+\lam^2)
+\frac{E^4(1-E^{4n})}{(1-E^2)^2}\,\lam^3,\\
b_\lam
&:=-\frac{E(1-E^{4n+2})}{(1-E^2)^2}\,(1-\lam)
+\frac{E^3(1-E^{4n})}{(1-E^2)^2}\,(\lam^2-\lam^3),\\
c_\lam
&:= -\frac{E(1-E^{4n+4})}{(1-E^2)^2}\,(1-\lam)
+\frac{E^3(1-E^{4n+2})}{(1-E^2)^2}\,(\lam^2-\lam^3),\\
d_\lam
&:= -\frac{E^{2n+1}}{1-E^2}(1-\lam),\\
e_\lam
&:=-\frac{E^{2n+2}}{1-E^2}\,(\lam-\lam^2),\\
f_\lam
&:=-\frac{E^{2n+3}}{1-E^2}\,(\lam^2-\lam^3).
\ea
\ep
\bprf
For all $e\in (-1,1)$, the  matrix 
$\Sigma_{\lam}(e-\i 0)\in\C^{4\times 4}$ has the structure
\ba
\Sigma_\lam(e-\i 0)
=\begin{bmatrix}
A_\lam(e) & B_\lam(e)\\
C_\lam(e) & D_\lam(e)
\end{bmatrix},
\ea
where the matrix-valued functions 
$A_\lam,\ldots ,D_\lam:(-1,1)\to\C^{2\times 2}$ are defined by 
\ba
A_\lam
&:=\frac{1}{1-E^2}
\begin{bmatrix}
1-\lam E^2 & (1-\lam) E \\
(1-\lam) E & 1-\lam E^2 
\end{bmatrix},\\
B_\lam
&:=(1-\lam)\, \frac{ E^{2n+1}}{1-E^2}
\begin{bmatrix}
1 & E\\
E & E^2
\end{bmatrix}\\
C_\lam
&:=\sigma_1 B_\lam\sigma_1,\\
D_\lam
&:=A_\lam.
\ea
A lengthy  calculation then leads to the assertion.
\eprf

In the following lemmas, we display some ingredients used in the proof of Theorem \ref{thm:nessc}. Recall from there that, for all $x\in\Z$ and 
$i,j=1,\ldots ,4$, the vector-valued functions 
$\xi_x, \eta_x: (-1,1)\to\C^4$ and the matrix-valued function 
$\Theta:(-1,1)\to\C^{4\times 4}$ are given by 
\ba
\xi_{x,i}
&={\big\langle\wt\delta^2_i,
\wt\rho_{1,+}^{}\wt\delta_x\big\rangle}_{\!2},\\ 
\eta_{x,i}
&=\varrho_{\delta^1_i,\delta_x}(\,\cdot-\i 0),\\
\Theta_{ij}
&={\big\langle\wt\delta^2_i,\wt\rho_{1,+}^{}\wt\delta^2_j\big
\rangle}_{\!2}.
\ea
The first lemma displays the explicit form of these functions.
\bl[Ingredients, 1]
\label{lem:ingr}
For $x\in\Z$, the functions 
$\xi_x, \eta_x: (-1,1)\to\C^4$ and $\Theta:(-1,1)\to\C^{4\times 4}$ read 
\ba
\label{xi-3}
\xi_x
&=\frac{E}{\pi\i (1-E^2)} \sum_{\sg=\pm}\rho_{\sg}
\big[E^{\sg(x+n+1)}, E^{\sg(x+n)},E^{\sg(x-n)},E^{\sg(x-n-1)}\big],\\
\label{eta-3}
\eta_x
&=-\frac{2E}{1-E^2}\,
\big[E^{|x+n|},
E^{|x+n+1|},
E^{|x-n-1|},
E^{|x-n|}\big],\\
\label{M-3}
\Theta
&=\frac{E}{\pi\i (1-E^2)}\sum_{\sg=\pm}\rho_{\sg}
\begin{bmatrix}
1 &  E^\sg &  E^{\sg(2n+1)} &  E^{\sg(2n+2)}\\
E^{-\sg} & 1 &  E^{\sg 2n} &  E^{\sg(2n+1)}\\
E^{-\sg(2n+1)} &  E^{-\sg 2n} & 1 &  E^\sg\\
E^{-\sg(2n+2)} &  E^{-\sg(2n+1)} & E^{-\sg} & 1 
\end{bmatrix},
\ea
where, for $\sg=\pm$, the function $\planck_{\sg}:\R\to\R$ is defined by
\ba
\planck_{\sg}(e)
:={\big(1+\e^{(\beta+\sg\delta) e}\big)}^{\!-1}.
\ea
Moreover, for all $e\in(-1,1)$, we have
\ba
\label{inv-eta}
\Sigma_\lam^{-1}(e-\i 0)\eta_x
=\frac{1}{\Delta_\lam(e)}
\big[a_{\lam,x}(e), b_{\lam,x}(e), c_{\lam,x}(e), d_{\lam,x}(e)\big],
\ea
where the component functions 
$a_{\lam,x},\ldots ,d_{\lam,x}:(-1,1)\to\C$ are given, 
for $x=n$,\footnote{Due to their complicated form, we do not display the formulas for general $x\in\Z$.} by  
\ba
a_{\lam,n}
&:=-\frac{2E^{2n+1}}{1-E^2},\\
b_{\lam,n}
&:=-\frac2{E^{2n+2}}{1-E^2}\,\lam,\\
c_{\lam,n}
&:=-\frac{2E^2(1-E^{4n+2})}{(1-E^2)^2}\,\lam
+\frac{2E^4(1-E^{4n})}{(1-E^2)^2}\,\lam^3,\\
d_{\lam,n}
&:=-\frac{2E(1-E^{4n+2})}{(1-E^2)^2}
+\frac{2E^3(1-E^{4n})}{(1-E^2)^2}\,\lam^2,
\ea
and, for $x=n+2$, by  
\ba
a_{\lam,n+2}
&:=-\frac{2E^{2n+3}}{1-E^2}\,\lam,\\
b_{\lam,n+2}
&:=-\frac{2E^{2n+4}}{1-E^2}\,\lam^2,\\
c_{\lam,n+2}
&:=-\frac{2E^2 (1-E^{4n+4})}{(1-E^2)^2}
+\frac{2E^4 (1-E^{4n+2})}{(1-E^2)^2}\,\lam^2,\\
d_{\lam,n+2}
&:=-\frac{2E^3 (1-E^{4n+2})}{(1-E^2)^2}\,\lam
+\frac{2E^5 (1-E^{4n})}{(1-E^2)^2}\,\lam^3.
\ea
\el
\bprf
Note that, for all $x\in\Z$, we have
\ba
\label{delta-3}
\wt\delta_x
=\left(\frac{E}{\pi\i(1-E^2)}\right)^{1/2}\, [E^x,E^{-x}],
\ea
and that the density $\rho_{1,+}\in\mL(\fh)$ of the XY NESS given in Theorem \ref{thm:ness} acts, for all $\eta\in\wt\fh$, as the matrix
multiplication operator 
\ba
\label{s-energy}
\wt\rho_{1,+}\eta
=\diag(\planck_+, \planck_-)\eta.
\ea
Hence, we get \eqref{xi-3} and \eqref{M-3}, and the expressions in
\eqref{eta-3} are given in Lemma \ref{lem:res}. Moreover, \eqref{inv-eta}
directly follows from Proposition \ref{prop:inv}.
\eprf

We next display some further ingredients used in the proof of Theorem 
\ref{thm:nessc}. Recall from there that, for $x=n$ and $y=n+2$, the function $F_\lam(x,y)=(w(h,h_\lam)\delta_x,\rho_{1,+}^{} w(h,h_\lam)\delta_y)$ reads
\ba
\label{Fl-2'}
F_\lam(n, n+2)
= F_1(n, n+2)
+\int_{-1}^1\rd e\,\,
\frac{\sum_{i=0}^8 p_i(e)\lambda ^i}{\sum_{i=0}^8 q_i(e)\lambda^{i}},
\ea
where the coefficient functions $p_i, q_i: (-1,1)\to\C$ with $i=1,\ldots ,8$ 
are displayed  in the following lemma.

\bl[Ingredients, 2]
\label{lem:coeff}
The numerator functions
$p_i: (-1,1)\to\C$ with $i=1,\ldots ,8$ have the form
\ba
p_0
&=\frac{(1-E^{4n+4})^2}{\pi\i E^{4n+1}(1-E^2)^5}\,
(\planck_+ E^4+\planck_-),\\
p_1
&=\frac{(1+E^2)(1-E^{4n+2})(1-E^{4n+4})}
{\pi\i E^{4n+1}(1-E^2)^3}\, \planck_-,\\
p_2
&=-\frac{2(1+E^2)(1-E^{4n+2})(1-E^{4n+4})}
{\pi\i E^{4n+1}(1-E^2)^5}\, (\planck_+ E^4+\planck_-),
\ea
\ba
p_3
&=-\frac{(1+E^{8n+4})(1+3E^2+E^4)-E^{4n}(1+E^2+5E^4+3E^6)}{\pi\i E^{4n+1}(1-E^2)^3}\,\planck_-\nonumber\\
&+\frac{E^3}{\pi\i (1-E^2)}\,\planck_+,\\
p_4
&= \frac{(1+E^{8n+4})(1+4E^2+E^4)
-E^{4n}(1+10E^4+E^8)}{\pi\i E^{4n+1}(1-E^2)^5}\, 
(\planck_+E^4+\planck_-),\\
p_5
&= \frac{2 E(1+E^2)(1-E^{4n})(1-E^{4n+2})}
{\pi\i E^{4n} (1-E^2)^3}\, \planck_-,\\
p_6
&= -\frac{2 E(1+E^2)(1-E^{4n})(1-E^{4n+2})}
{\pi\i E^{4n} (1-E^2)^5}\, (\planck_+E^4+\planck_-),\\
p_7
&= -\frac{E^3(1-E^{4n})^2}{\pi\i E^{4n}(1-E^2)^3}\,
\planck_-,\\
p_8
&= \frac{ E^3(1-E^{4n})^2}{\pi\i E^{4n}(1-E^2)^5}\,
(\planck_+ E^4+\planck_-),
\ea
and the denominator functions $q_i: (-1,1)\to\C$ with $i=1,\ldots ,8$ read $q_1=q_3=q_5=q_7=0$, 
\ba
q_0
&=-\frac{(1-E^{4n+4})^2}{E^{4n}(1-E^2)^4},\\
q_2
&=\frac{2(1+E^2)(1-E^{4n+2})(1-E^{4n+4})}{E^{4n}(1-E^2)^4},
\\
q_4
&=-\frac{(1+E^4)(1-E^{4n})(1-E^{4n+4})+4E^2(1-E^{4n+2})^2}
{E^{4n}(1-E^2)^4},\\
q_6
&=\frac{2E^2(1+E^2)(1-E^{4n})(1-E^{4n+2})}{E^{4n}(1-E^2)^4},
\\
q_8
&=-\frac{E^4(1-E^{4n})^2}{E^{4n}(1-E^2)^4}.
\ea
Moreover, on the energies $e=\eps(k)$ with $k\in[0,\pi]$, we have 
$\Im[p_{2i+1}(\eps(k))]=0$ for $i=0,2,3$ and, for $i=1$, we find
\ba
\Im[p_3(\eps(k))]
=-\frac{\eps(k)}{\pi}\, [\planck_{\beta_L}(\eps(k))
-\planck_{\beta_R}(\eps(k))].
\ea
Moreover, on these energies, the denominator functions have the form
\ba
\label{q0cos}
q_0(\eps(k))
&=\frac{\sin^2(2(n+1)k)}{4\sin^4(k)},\\
\label{q2cos}
q_2(\eps(k))
&=-\frac{\sin((2n+1)k)\sin(2(n+1)k)\cos(k)}{\sin^4(k)},
\ea
\ba
\label{q4cos}
q_4(\eps(k))
&=\frac{\sin(2nk)\sin(2(n+1)k)\cos(2k)+2\sin^2((2n+1)k)}{2\sin^4(k)},\\
\label{q6cos}
q_6(\eps(k))
&=-\frac{\sin((2n+1)k)\sin(2nk)\cos(k)}{\sin^4(k)},\\
\label{q8cos}
q_8(\eps(k))
&=\frac{\sin^2(2nk)}{4\sin^4(k)}.
\ea
\el
\bprf
This follows from Lemma \ref{lem:ingr} and a lengthy calculation.
\eprf

Finally, in the following lemma, we discuss the ingredients needed in order to derive the absolute convergence of the NESS current integral in the proof of Theorem \ref{thm:nessc}. Recall from there that the function
$Q_\lam: (-1,1)\to\R$ is given by 
\ba
Q_\lam
=|\Delta_\lam|^2
=|\!\det(\Sigma_\lam(\,\cdot-\i0)|^2
=\sum_{i=0}^4 q_{2i}\,\lambda^{2i}.
\ea
This function has the following property.
\bl[Boundedness]
\label{lem:roots}
For $\lam\in\R\setminus\{0\}$, the inverse of 
$Q_\lam$ is bounded. In particular, for $\lam=\pm 1$, we have
$Q_{\pm 1}=1$.
\el
\bprf
For $\lam=1$, we have $\Sigma_1(e-\i0)= 1\in\C^{4\times 4}$ for 
all $e\in [-1,1]$ from \eqref{mat-0}. Since the determinant \eqref{Delta-1} contains even powers of $\lam$ only, we thus have $Q_{\pm 1}=1$ on $[-1,1]$. We next observe that
\ba
\label{Qinv}
\frac{1}{Q_\lam}
=-\frac{E^{4n} (1-E^2)^4}{\prod_{i=1}^4P_{\lam,i}},
\ea
where the polynomials $P_{\lam,i}:(-1,1)\to\C$ with $i=1,\ldots ,4$ are
defined by 
\ba
\label{P1}
P_{\lam,1}
&:= 1+E^{2n+2}-(1+E^{2n})\lam^2,\\
\label{P2}
P_{\lam,2}
&:= -1+E^{2n+2}+(1-E^{2n})\lam^2,\\
\label{P3}
P_{\lam,3}
&:=1-E^{2n+2}-E^2(1-E^{2n})\lam^2,\\
\label{P4}
P_{\lam,4}
&:=-1-E^{2n+2}+ E^2(1+E^{2n})\lam^2.
\ea
From now on, let $\lam\in\R\setminus\{0,\pm 1\}$. Then, it easily
follows from \eqref{P1}--\eqref{P4} that, for all $n\in\N_0$, there are no unimodular roots of $P_{\lam,1}$ and $P_{\lam,4}$, whereas the only 
unimodular roots of  $P_{\lam,2}$ and $P_{\lam,3}$ are $E=\pm 1$. We
next study the order of these roots.
Specializing \eqref{P1}--\eqref{P4} for $n=0$, we see that, in this case,
 the roots are simple and we  get
\ba
\frac{1}{Q_\lam}
=-\frac{(1-E^2)^2}{(E^2+1-2\lam^2)(1+(1-2\lam^2)E^2)},
\ea
which implies the assertion for $n=0$. Next, let $n>0$ and let us first consider $P_{\lam,2}$ and $E=1$. From the factorization
$P_{\lam,2}=(E-1) R_{\lam,2}$, where we set
\ba
R_{\lam,2}
:=E^{2n} (1+E)+(1-\lam^2)\sum_{i=0}^{2n-1}E^i,
\ea
we get $R_{\lam,2}(1)=2(n+1-n\lam^2)$. Hence, if 
$\lam^2\neq (n+1)/n$,
the polynomial $P_{\lam,2}$ has a simple root at $E=1$. On the other hand, if $\lam^2=(n+1)/n$, we can write $P_{\lam,2}=(E-1)^2 S_{2}$,
where 
\ba
S_2
:=E^{2n}+\frac1n \sum_{i=0}^{2n-1} (2n-i) E^{2n-1-i},
\ea
and now we have $S_2(1)=2(n+1)\neq 0$. Hence, in this case,
$P_{\lam,2}$ has a double root at $E=1$. Since 
$P_{\lam,2}(-E)=P_{\lam,2}(E)$, the same conclusions hold for the root
$E=-1$. Moreover, $P_{\lam,3}$ can be treated similarly (we again have the two cases $\lam^2$ different or equal to $(n+1)/n$) and the conclusions remain unchanged. Hence, the order of the roots
$E=\pm 1$ in the denominator of \eqref{Qinv} is not exceeding $4$
and, since the numerator cancels these singularities of 
$1/Q_\lam$, we arrive at the assertion.
\eprf
\section{Van Hove weak coupling theory}
\label{app:davies}

The material of the following theorem is taken from Davies 
\cite{Davies1,Davies2}. We do not display his assertions in full
generality but rather adapt them to our special case at hand. The 
ingredients are as follows.
Let $\mH$ be a Hilbert space, $P_0\in\mL(\mH)$ a  projection, and 
$P_1:=1-P_0\in\mL(\mH)$. Let $U^t\in\mL(\mH)$ with $t\in\R$ be a strongly
continuous one-parameter group of isometries s.t., for all $t\in\R$, we
have
\ba
\label{UP0}
[U^t,P_0]=0.
\ea
The generator of $U^t$ is denoted by $Z$. Moreover, let 
$A\in\mL(\mH)$ with $A^\ast=-A$, and let $V_\lam^t\in\mL(\mH)$ with $\lam,t\in\R$ be the one-parameter group generated by $Z+\lam A$.
Besides, the operators $W_\lam^t, R_\lam^t\in\mL(\mH)$ 
with $\lam,t\in\R$ are defined by
\ba
\label{W}
W_\lam^t
&:=P_0 V_\lam^t P_0,\\
\label{R}
R_\lam^t
&:= P_1 V_\lam^t P_0.
\ea
Moreover, if $\dim(\ran(P_0))<\infty$, we define the spectral average
$X^\natural\in\mL(\mH)$ of an operator $X\in\mL(\mH)$ by
\ba
\label{Xnat}
X^\natural
:=\lim_{T\to\infty}\frac{1}{2T}\int_{-T}^T\!\!\rd t\,\,U^tP_0XP_0U^{-t}.
\ea

We then have the following result.
\bt[Van Hove weak coupling limit]
\label{thm:davies}
Let us assume the validity of the conditions
\begin{enumerate}
\item[(a)] $\dim(\ran(P_0))<\infty$,
\item[(b)]  $P_0AP_0=0$ and $P_1AP_1=0$,
\item[(c)] $\int_0^\infty\!\!\rd t\,\,\|P_0AP_1 U^t P_1 A P_0\|<\infty$.
\end{enumerate}
Then, the following assertions hold.
\begin{enumerate}
\item[(1)] For $t_0>0$ and all $\psi\in\mH$, we have
\ba
\label{davies-1}
\lim_{\lam\to 0}\, \sup_{t\in[0,t_0]}
\big\|U^{-t/\lam^2}W_\lam^{t/\lam^2}P_0\psi-\e^{tK^\natural}P_0\psi\big\|
=0,
\ea
where $K^\natural\in\mL(\mH)$ is the spectral average of $K\in\mL(\mH)$
given by
\ba
\label{K}
K
:=\int_0^\infty\!\!\rd t\,\, U^{-t} P_0AP_1 U^t P_1 A P_0.
\ea
\item[(2)] Let $B\in\mL(\mH)$ satisfy $B\ge 0$ and $[B,U^t]=0$ for
all $t\in\R$,  and let the operator $S_B\in\mL(\mH)$ be defined by
\ba
\label{SB}
S_B
:=-2 \int_0^\infty\!\!\rd t\,\, \Re[P_0AP_1 BU^t P_1AP_0 U^{-t}],
\ea
where the integral is assumed to converge in norm. For
$t_0>0$ and all $\varphi,\psi\in\mH$, we have
\ba
\label{davies-2}
\lim_{\lam\to 0}\, \sup_{t\in[0,t_0]}
\left|(R_\lam^{t/\lam^2}P_0\varphi, B R_\lam^{t/\lam^2}P_0\psi)-\int_0^t\!\!\rd s\,\,
(\e^{sK^\natural} P_0\varphi, S_B^\natural \e^{sK^\natural} P_0\psi)\right|
=0.
\ea
\end{enumerate}
\et
\bprf
See Davies \cite{Davies1,Davies2}.
\eprf

Recall from \eqref{psia} that, for 
$a\in\{L,R\}$ and $\beta\in\R$, the function $\psi_a^\beta:\R\to\C$ is
given by
\ba
\psi_a^\beta(t)
=(\delta_{\mR,a},\ia\planck_\beta(h_a)\e^{\i th_a}\ias \delta_{\mR,a}).
\ea
We then have the following lemma which is used in the proof of Theorem
\ref{thm:ma-ness}.
As usual, in the proofs below, the constant $C$
can take different values at each place it appears. Moreover, the Laplace
transform at the points $\i\eps$ with $\eps\in\R$ is denoted by $\Psi^\beta_a(\i\eps):=\int_0^\infty\rd t\,\, \psi^\beta_a(t)\,\e^{-\i\eps t}$
and the temporal Fourier transform by 
$\hat\psi^\beta_a(\eps):=\int_{-\infty}^\infty\rd t\,\, \psi^\beta_a(t)\,\e^{-\i\eps t}$.
\bl[Reservoir time correlation]
\label{lem:rtc}
For $a\in\{L,R\}$ and $\beta\in\R$, there exists a constant 
$C>0$ s.t., for all $t\in\R$, we have 
\ba
\label{rtc-1}
\big|\psi_a^\beta(t)\big|
\le C(1+|t|)^{-3/2}.
\ea
Moreover, for all $\eps\in\R$, it holds that
\ba
\label{rtc-2}
\Re[\Psi^\beta_a(\i\eps)]
=\frac12\,\hat \psi^\beta_a(\eps)
=\begin{cases}
2(1-\eps^2)^{1/2}\planck_\beta(\eps), & |\eps|< 1\\
0, & |\eps|\ge 1.
\end{cases}
\ea
\el
\bprf
In order to prove the first assertion, we switch to the energy space representation of the subreservoir Hamiltonians $h_a$ from Lemma \ref{lem:subres} already used in 
the derivation of \eqref{psia-1} in order to make the decomposition
$\psi_a^\beta=\frac2\pi(\varphi_\beta+\bar\varphi_{-\beta})$, 
where the function $\varphi_\beta:\R\to\C$, independent of 
$a\in\{L,R\}$, is defined by
\ba
\varphi_\beta(t)
:=\int_0^1\rd e\,\,(1-e^2)^{1/2}\planck_\beta(e)\,\e^{\i te}.
\ea
After the coordinate transformation $e\mapsto 1-e$ and one partial integration, we get 
\ba
\varphi_\beta(t)
=\frac{\i}{2t}+\frac{\e^{\i t}}{\i t}\sum_{i=1}^3 \varphi_{\beta,i}(t),
\ea
where the functions $\varphi_{\beta,i}:\R\to\C$ have the form
$\varphi_{\beta,i}(t)
:=\int_0^1\rd e\,\,\phi_{\beta,i}(e)\, \e^{-\i te}$,
the integrands 
$\phi_{\beta,i}:(0,1)\to[0,\infty)$ are defined by
\ba
\phi_{\beta,1}(e)
&:=\frac12\, e^{-1/2} (2-e)^{1/2}\planck_\beta(1-e),\\
\phi_{\beta,2}(e)
&:=\frac12\, e^{1/2}(2-e)^{-1/2}\planck_\beta(1-e),\\
\phi_{\beta,3}(e)
&:=\beta\,e^{1/2}(2-e)^{1/2}\planck_\beta(1-e)
\planck_{-\beta}(1-e),
\ea
and we used the fact that $\varrho'_\beta(e)=-\beta \planck_\beta(e)\planck_{-\beta}(e)$ (the prime denotes the derivative w.r.t. $e$). 
First, using $|\phi_{\beta,1}(e)|\le C e^{-1/2}$ for all $e\in (0,1)$,
we immediately get $|\varphi_{\beta,1}(t)|\le C$ for all $0\le|t|\le 1$.
On the other hand, if $|t|>1$, we write 
\ba
\varphi_{\beta,1}(t)
=\int_0^{1/|t|}\rd e\,\,\phi_{\beta,1}(e)\,\e^{-\i te}
+\int_{1/|t|}^1\rd e\,\,\phi_{\beta,1}(e)\,\e^{-\i te}.
\ea
Using directly $|\phi_{\beta,1}(e)|\le C e^{-1/2}$ for all $e\in (0,1)$ for the first integral and one partial integration 
and $|\phi'_{\beta,1}(e)|\le C e^{-3/2}$ for all $e\in (0,1)$ for the
second integral, we can bound the modulus of both integrals by
 $C|t|^{-1/2}$ for $|t|>1$.
Moreover, for $i=2,3$, since 
$|\phi_{\beta,i}(e)|\le C$ for all $e\in (0,1)$, we have
 $|\varphi_{\beta,i}(t)|\le C$ for all $0\le|t|\le 1$.
If $|t|>1$, using $|\phi'_{\beta,i}(e)|\le C\e^{-1/2}$ for all 
$e\in (0,1)$ and one partial integration, we get
$|\varphi_{\beta,i}(t)|\le C|t|^{-1}$.
These facts imply \eqref{rtc-1}. 
We next turn to the proof of the second assertion. The real part of the Laplace transform (which exists due to \eqref{rtc-1}) can be written as
\ba
\label{Re-1}
\Re[\Psi^\beta_a(\i\eps)]
&=\frac12 \int_{-\infty}^\infty\rd t\,\, \psi^\beta_a(t)\,
\e^{-\i t\eps}\\
&=\lim_{\delta\to 0^+} \frac{1}{\pi}
\int_{-\infty}^\infty\rd t\,\,\e^{-\delta |t|}\e^{-\i t\eps}\int_{-1}^1\rd e\,\, \xi_\beta(e)\, \e^{\i t e},
\ea
where, in the first equality, we used that $\psi^\beta_a(-t)=\bar\psi^\beta_a(t)$ for all $t\in\R$ (leading to a Fourier transform), and,  in the second, we set $\xi_\beta(e):= (1-e^2)^{1/2}\planck_\beta(e)$ and we used \eqref{psia-1} and Lebesgue's theorem
of dominated convergence. Since
$\int_{-\infty}^\infty\rd t\,\, \e^{-\delta |t|}\e^{\i t (e-\eps)}
=2\delta/(\delta^2+(e-\eps)^2)$, using Fubini's theorem, and transforming the coordinates as
$e\mapsto (e-\eps)/\delta$, we get
\ba
\Re[\Psi^\beta_a(\i\eps)]
=\lim_{\delta\to 0^+} \frac{2}{\pi}
\int_{-(1+\eps)/\delta}^{(1-\eps)/\delta}\rd e\,\,
\frac{\xi_\beta(\eps+\delta e)}{1+e^2}.
\ea
If $|\eps|>1$, using $0\le \xi_\beta(e)\le 1$ for all $e\in[-1,1]$, we
directly get $\Re[\Psi^\beta_a(\i\eps)]=0$. For $|\eps|\le1$, we decompose
the numerator as $\xi_\beta(\eps+\delta e)=\xi_\beta(\eps)+[\xi_\beta(\eps+\delta e)-\xi_\beta(\eps)]$. Due to 
$|\xi_\beta(\eps+\delta e)-\xi_\beta(\eps)|\le C\delta |e|$
for all $e$ in the integration interval, 
the difference term vanishes and we get
$\Re[\Psi^\beta_a(\i\eps)]=2 \xi_\beta(\eps)$. Since $\xi_\beta(\pm 1)=0$, we arrive at 
\eqref{rtc-2}.
\eprf
\end{appendix}

\end{document}